\documentclass[%
 reprint,
nofootinbib,
 amsmath,amssymb,
 aps,
]{revtex4-2}

\pdfoutput=1
\usepackage[pdftex]{graphicx}
\usepackage[pdftex]{color}
\usepackage[colorlinks,bookmarks]{hyperref}
\definecolor{linkblue}{rgb}{0,0,0.8}
\definecolor{linkgreen}{rgb}{0,0.5,0}
\hypersetup{linkcolor=linkblue, citecolor=linkgreen, urlcolor=linkblue}

\usepackage{dcolumn}
\usepackage{bm}
\usepackage{xcolor} 
\usepackage{hyperref}

\usepackage[normalem]{ulem}

\newcommand\aap{Astron.~Astrophys.}
\newcommand\apjl{Astrophys.~J.~Lett.}
\newcommand\apjs{Astrophys.~J.~Suppl.~Ser.}
\newcommand\mnras{Mon.~Not.~R.~Astron.~Soc.}
\newcommand\jcap{J.~Cosmo.~Astropart.~Phys.}
\newcommand\physrep{Phys.~Rep.} 
\newcommand\apss{Astrophy.~Space~Sci.} 
\newcommand\pasp{Publ.~Astron.~Soc.~Pac.} 

\newcommand{\pd}{\partial}
\newcommand{\bs}{\boldsymbol}

\definecolor{valecol}{rgb}{0,0.5, 1.}
\definecolor{davicol}{rgb}{0.2,0.6, 1.}
\definecolor{henricol}{rgb}{0.2,0.3, 1.}

\begin{document}

\preprint{...}

\title{Revisiting constraints on asymmetric dark matter from collapse in white dwarf stars}

\author{Heinrich Steigerwald$^{1}$}
 \email{heinrich@steigerwald.name}
\author{Valerio Marra$^{1,2,3,4}$}%
 \email{valerio.marra@me.com}
\author{Stefano Profumo$^{5}$}
 \email{profumo@ucsc.edu}
\affiliation{$^{1}$Núcleo de Astrofísica e Cosmologia, Universidade Federal do Espírito Santo, 29075-910, Vitória, ES, Brazil\\
$^{2}$Departamento de Física, Universidade Federal do Espírito Santo, 29075-910, Vitória, ES, Brazil\\
$^{3}$INAF -- Osservatorio Astronomico di Trieste, via Tiepolo 11, 34131, Trieste, Italy\\
$^{4}$IFPU -- Institute for Fundamental Physics of the Universe, via Beirut 2, 34151, Trieste, Italy\\
$^{5}$Department of Physics and Santa Cruz Institute for Particle Physics, 1156 High Street, University of California, Santa Cruz, California 95064, USA}

\date{\today}

\begin{abstract}
The runaway collapse phase of a small dark matter cluster inside a white dwarf star encompasses a reversible stage, where heat can be transferred back and forth between nuclear and dark matter. Induced nuclear burning phases are stable and early carbon depletion undermines previous claims of type Ia supernova ignition.
Instead, mini black holes are formed at the center of the star that either evaporate or accrete stellar material until a macroscopic sub-Chandrasekhar-mass black hole is formed. 
In the latter case, a 0.1 to 1 second lasting electromagnetic transient signal can be detected upon ejection of the white dwarf's potential magnetic field. 
Binary systems that transmute to black holes and subsequently merge emit gravitational waves. Advanced LIGO/Virgo should detect one such sub-Chandrasekhar binary black hole inspiral per year, while future Einstein telescope-like facilities will detect thousands per year.
The effective spin parameter distribution is peaked at 0.2 and permits future studies to disentangle from primordial sub-Chandrasekhar black holes. 
Such signatures are compatible with current direct detection constraints, as well as with neutron star constraints in the case of bosonic dark matter, even though they remain in conflict with the fermionic case for part of the parameter space.
\end{abstract}

\maketitle





\section{Introduction}\label{sec:introduction}

Dark matter (DM) collapse under self-gravity has first been studied by \citet{1989PhRvD..40.3221G} in the context of neutron stars (NSs). The authors show that a critical number $N_{\rm sg}$ of particles has to  accumulate for collapse under self-gravity to initiate. This criterion was thence popularized as the ``collapse criterion" and applied to explore DM phenomenology in NSs \cite{2011PhRvD..83h3512K,2015PhRvD..91k5001B,2018PhRvL.121v1102K,2021PhRvL.126n1105D}, white dwarf (WD) stars \cite{2011PhRvD..83h3512K,Bramante:2015cua,2018PhRvD..98k5027G,2019PhRvD.100c5008J,2019PhRvD.100d3020A}, and main sequence stars \cite{2016PhRvD..93b3508K,2021MNRAS.507.3434R}. Yet, little attention has been devoted to the dynamics of collapse inside the star itself. 

The process is intrinsically iterative: scattering between DM and stellar matter (SM) particles reduces the total energy of DM particles, while gravitational self-attraction and the resulting orbital hardening increases the DM kinetic energy. The runaway nature of the process lasts as long as heat imparted to the scattered SM particles is efficiently evacuated to the rest of the star. The process is also reversible, as long as scatterings are elastic. If SM particles became more energetic, on average, than DM particles, the process would reverse and the DM cluster would expand.

The question of heat evacuation in the baryonic component has been addressed for NSs \citep{1989PhRvD..40.3221G}, but so far, to our knowledge, never in the context of nondegenerate stars (including WD stars, which are electron-degenerate, but not nucleon-degenerate). 

Here, we seek an answer by deriving a system of first-order differential equations that permit us to follow DM and nuclear macroscopic properties along the elastic collapse phase.
We are particularly interested in previously investigated astrophysical phenomenology, including claims of type Ia supernova (SN Ia) ignition \citep{Bramante:2015cua,2018PhRvD..98k5027G,2019PhRvD.100c5008J,2019PhRvD.100d3020A} 
and collapse to a black hole (BH)  \citep{2011PhRvD..83h3512K,2019PhRvD.100d3020A}.

The present-day understanding is that ``normal" SNe Ia originate either from deflagrations with transition to detonation in Chandrasekhar-mass ($\sim 1.4$~$M_{\odot}$) WDs \citep{1991A&A...245..114K,1995ApJ...443...89H}, or from pure detonations in sub-Chandrasekhar-mass WDs \cite{1969Ap&SS...5..180A,2010ApJ...714L..52S,2017MNRAS.470..157B,2018MNRAS.474.3931B,2021ApJ...909L..18S}. While the former channel unlikely produces all events \cite{2010Natur.463..924G}, the latter is still lacking a convincing ignition mechanism \citep{2021MNRAS.503.4734P}, see, however, Ref.~\cite{2021PhRvL.127a1101S}. 
Additionally, pure deflagrations can reproduce certain types of ``peculiar" SNe Ia \citep{1976Ap&SS..39L..37N,2003PASP..115..453L,2012ApJ...761L..23J,2014MNRAS.438.1762F}, and their ignition from DM collapse has been studied from thermonuclear ignition \cite{Bramante:2015cua,2018PhRvD..98k5027G,2019PhRvD.100c5008J,2019PhRvD.100d3020A}, pycnonuclear ignition \cite{2020PhRvD.102h3031H}, and Hawking radiation ignition \cite{2018PhRvD..98k5027G,2019PhRvD.100d3020A}. It has also been questioned if the observed correlation between SN Ia magnitudes and host galaxy masses has its origin in the local DM environment \citep{2022MNRAS.510.4779S}.

The formation of a mini BH inside a WD can lead to the implosion of the latter. In principle, this can generate BHs of mass 0.3--1.4 $M_\odot$, which may interact with other BHs, NSs, or WDs, and generate detectable gravitational waves (GWs)  \citep{2021PhRvL.126n1105D}. While generally, the observation of a BH with mass $< 1.4$~$M_{\odot}$ is considered a smoking gun of exotic new physics \citep{1975Natur.253..251C},
BHs in this mass range could be attributed to either DM collapse inside a NS \citep{2018PhRvL.121v1102K}, to a primordial BH from the QCD phase transition (produced with 0.7~$M_{\odot}$ and taking into account accretion of gas \cite{1982Natur.298..538C,1997PhRvD..55.5871J,2017PTEP.2017h3E01C,2020ARNPS..70..355C}), to the capture of a primordial BH with asteroid mass by a NS and subsequent transmutation to a macroscopic BH (\cite{2008PhRvD..78c5009G}, even though constraints on these primordial BHs \cite{2013PhRvD..87l3524C,2014JCAP...06..026P}  were largely overestimated  \cite{2020PhRvD.102h3004G,2019JCAP...08..031M}), or to BH formation through atomic DM \citep{2018PhRvL.120x1102S}.
As we shall see, given that WDs account for the final evolutionary state of 97\% of main sequence stars \citep{2001PASP..113..409F},
GWs from the coalescence of binary BHs from transmuted binary WDs should already be observed with the advanced Laser Interferometer Gravitational-Wave Observatory (aLIGO) and Virgo Interferometer \cite{2018PhRvL.121w1103A,2018LRR....21....3A}.
In addition, the effective inspiral spin parameter distribution is inherited from the progenitor WD angular momentum distribution and permits to disentangle this sub-Chandrasekhar BH formation channel from others.

The present work investigates collapse of nonself-annihilating DM, such as, for example, asymmetric DM \citep{1985PhLB..165...55N,1989PhRvD..40.3221G,2011PhRvD..83h3512K,2013IJMPA..2830028P,2014PhR...537...91Z}. 
Asymmetric DM is motivated as it can explain the matter-antimatter asymmetry in the Universe (see, e.g., Ref.~\cite{2013IJMPA..2830028P}, for a review). 
For the sake of generality, we also include a Yukawa-type nongravitational attractive self-interaction with potential $V(r) = \alpha\exp(-\mu r)/r$, where $\mu$ is
the mediator mass and $\alpha$ a coupling constant.

We use natural units, $c=k_{\rm B}=\hbar=1$, while keeping $G$ explicit. 
Stellar quantities are indexed with an asterisk (*), while DM quantities are left without. The infinity symbol ($\infty$) indicates stellar core quantities far from the DM cluster. We use the term stellar matter (SM) to designate initially ions, but later, as collapse proceeds and locally heats the center of the star, these are crushed to nucleons and then to partons (quarks and gluons). Therefore, we stick to the generic term SM particles keeping its meaning in mind.

The present paper is organized as follows.
In Sec.~\ref{sec:general} we derive the general set of ``elastic" collapse equations. In Sec.~\ref{sec:WD} we explore phenomenology in WD stars. 
We present our conclusions in Sec.~\ref{sec:conclusion}.

\section{General equations}\label{sec:general}

\subsection{DM capture and accumulation}\label{sec:accumulation}
Let $N(t)$ be the number of particles of a DM cluster at the center of a star with age $t$.
The number of captured particles during an interval $dt$ is $dt\;\!\Gamma_{\rm cap}$, where $\Gamma_{\rm cap}$ is the capture rate \citep{2017PhRvD..96f3002B,2020PhRvD.102d8301I}\footnote{see also Refs.~\cite{2019JCAP...08..018D,2021JCAP...11..056A} for some improvements and Appendix~A of Ref.~\cite{2021PhRvD.104l3031I} for derivations of analytical approximations.}
\begin{align}\label{eq:Gamma-capture-multi-text}
    \Gamma_{{\rm cap}} = &\; \frac{\sqrt{6\;\!\pi}\;\!R_*^2\;\!v_{\rm esc}^2\;\!\rho_{\rm gal}}{m\;\! v_{\rm gal}} \nonumber \\
    \times &\; \sum_{j=1}^{\infty}p_j(\tau)\;\!\Big[1+\delta - (\gamma_j+\delta)\;\!e^{-(\gamma_j-1)/\delta}\Big] \,,
\end{align}
where $v_{\rm esc}\equiv (2\;\!G\;\!M_*/R_*)^{1/2}$ is the escape velocity of the star, $R_*$ and $M_*$ are radius and mass of the star, respectively, $\delta \equiv 2\;\!v_{\rm gal}^2/3\;\!v_{\rm esc}^2$, $\gamma_j \equiv (1-\beta_+/2)^{-j}$, 
$\beta_{\pm} \equiv 4\;\! m m_*/(m\pm m_*)^2$, $m$ and $m_*$ are DM and SM particle masses, $\rho_{\rm gal}$ and $v_{\rm gal}$ are the galactic DM density and velocity dispersion, respectively, and 
\begin{align}\label{eq:Poisson-weighting-text}
    p_j(\tau) =&\; 2 \int_0^1 \frac{y e^{-y \tau}(y\tau)^j\;\!dy}{j!}
\end{align}
is a Poisson weighting that gives the probability of $j$ scatters for the optical depth $\tau \equiv 3 \sigma_* /2\;\!\sigma_{\rm sat}$, where $y$ is a kinematical quantity,  $\sigma_*$ is the DM-SM scattering cross section and $\sigma_{\rm sat}=R_*^2/N_*$ is the saturation cross section, where $N_*=M_*/m_*$ is the total number of SM particles. 
Once captured, DM particles settle at the center of the star, where they thermalize after a timescale (see Appendix~\ref{sec:thermalization} for a derivation)
\begin{align}\label{eq:t-th}
    t_{\rm th} =  \frac{3\;\!m}{\rho_*\;\!\sigma_*\;\!v_*}\Big[\frac{3\sqrt{2}\pi}{16}\frac{v_*}{v_{\rm gal}} + \frac{1}{2} + \ln\Big(\frac{m}{m_*}\Big)\Big]\,,
\end{align}
where $\rho_*$ is the central density of the star and $v_*$ the mean velocity of SM particles.
The number of captured and thermalized particles during an interval $dt$ is $dN$. Since $dt\;\!\Gamma_{\rm cap}$  must be equal to $(dt+dt_{\rm th})\;\!(dN/dt)$,
where $dt_{\rm th} \equiv t_{\rm th}(t+dt)-t_{\rm th}(t)$ is the increase of thermalization time during $dt$, the number increase rate of DM particles is
\begin{align}\label{eq:dNdt}
    \frac{dN}{dt} =&\; \Gamma_{\rm cap}\Big(1+\frac{dt_{\rm th}}{dt}\Big)^{-1}\,, & (t\geq t_{\rm th,0})
\end{align}
where $t_{\rm th,0}$ is the initial thermalization time, $dN/dt=0$ for $t<t_{\rm th,0}$, and $t_{\rm th,0}$ is the larger solution of $t = t_{\rm th}[T_{*\infty}(t)]$. 
The time derivative of the thermalization time is
\begin{align}
    \frac{dt_{\rm th}}{dt} =- \frac{3\;\!m}{2\;\!\rho_*\;\!\sigma_*\;\!v_*\;\!T_{*\infty}}\Big[\frac{1}{2}+\ln\Big(\frac{m}{m_*}\Big)\Big]\;\!\frac{dT_{*\infty}}{dt} \,,
\end{align}
where $T_{*\infty}$ is the stellar core background temperature and $dT_{*\infty}/dt$ its time derivative. If the stellar core temperature is constant, we have $dN/dt=\Gamma_{\rm cap}$ and, integrating, $N(t)=\Gamma_{\rm cap}\;\!t$.

\subsection{Self-attraction and collapse}\label{sec:collapse}
The mean potential energy per DM particle at the center of a star at $r=0$ is (see, e.g., Ref.~\citep{2018PhRvL.121v1102K} and Appendix~\ref{sec:potential-energy})
\begin{align}\label{eq:U}
    U =&\; - \frac{4\;\! \pi \;\! G \rho_*\;\! m\;\! R^2}{5} - \frac{3\;\! G N m^2}{5\;\! R} \nonumber\\
    &\;- \frac{3\alpha N e^{-\mu R_0}}{2\mu^2R^3}\big(3 + 3\mu R_0 + \mu^2 {R_0}^2 \big) \,,
\end{align}
where $R$ is the radius of the DM cluster and $R_0= R(4\pi/3N)^{1/3}$ is the mean interparticle distance. The terms on the right-hand side of Eq.~\eqref{eq:U} account for contributions coming from (1) gravitational attraction due to the density of the star, (2) gravitational self-attraction, and potentially (3) nongravitational self-attraction\footnote{For simplicity, we have assumed a constant distribution of DM particles inside $R$ (a step function), the differences with a Maxwell-Boltzmann distribution are minor (see Appendix~\ref{sec:potential-energy}). In Ref.~\cite{2018PhRvL.121v1102K}, the numerical factor in term (1) is $8\pi$ instead of $4\pi$.}.
It is easy to verify that the accumulation timescale $(dN/dt)^{-1}$ is much longer than the orbital timescale of DM particles $(R^3/GM)^{1/2}$, where $M=Nm$ is the total thermalized DM mass, hence the DM cluster energy repartition is given by the virial theorem, with the mean kinetic energy per particle $K = -U/2$, and the mean total energy per particle is $E = K + U = U/2 = - K$. 

Despite the sporadic nature of individual scatterings, the average \textit{per particle} energies can be treated as differential functions, and we can write, for example, $E$ as a total differential of $N$ and $R$,
\begin{align}\label{eq:dE}
    dE = \frac{\pd E}{\pd N}\;\!dN+ \frac{\pd E}{\pd R}\;\!dR \,,
\end{align}
where from Eq.~\eqref{eq:U} and using the virial theorem, 
\begin{align}
    \frac{\pd E}{\pd R} =&\; - \frac{4\pi G \rho_* m R}{5} + \frac{3 G N m^2}{10 R^2} + \frac{3\alpha N}{4R^4 \mu^2}f(\mu R_0)\,, \label{eq:dEdR}\\
    \frac{\pd E}{\pd N} =&\; - \frac{3 G m^2}{10 R}- \frac{\alpha}{4 R^3 \mu^2}f(\mu R_0) \,,\label{eq:dEdN}
\end{align}
where we have defined $f(y)\equiv e^{-y}(9+9y+4y^2+y^3)$.
As long as $\pd E/\pd R>0$, DM particles stay in thermal equilibrium with the star ($K=K_*$) and settle roughly inside the ``thermal radius" \citep{1989PhRvD..40.3221G}\footnote{Note that the numerical factor on the right-hand side of Eq.~\eqref{eq:Rth} is $9/8\pi$ in Ref.~\cite{1989PhRvD..40.3221G}, $15/8\pi$ in Ref.~\cite{2018PhRvL.121v1102K}, while it is $9/4\pi$ in Ref.~\cite{Bramante:2015cua}.}
\begin{align}\label{eq:Rth}
    R_{\rm th}  = \Big(\frac{15\;\! T_{*\infty}}{4\;\! \pi\;\! G \rho_*\;\! m}\Big)^{\!1/2}\,.
\end{align}
The actual radius during the adiabatic phase decreases slightly over time and has to be computed numerically in the general case, but in the absence of nongravitational self-attraction (i.e. $\alpha =0$), it is the larger positive solution of the cubic equation \eqref{eq:U}
\begin{align}\label{eq:R-adiabatic}
\frac{R}{R_{\rm th}} =&\left[- \frac{\eta}{2}+\Big(\frac{\eta^2}{2}-\frac{1}{27}\Big)^{1/2}\right]^{1/3} \nonumber \\
&+\frac{1}{3}\left[- \frac{\eta}{2}+\Big(\frac{\eta^2}{2}-\frac{1}{27}\Big)^{1/2}\right]^{-1/3} \,,
\end{align}
where $\eta \equiv N/N_{\rm sg}$, and where $N_{\rm sg}\equiv 4\pi\rho_*{R_{\rm th}}^3/3m$ is the critical number for self-gravitation of Ref.~\cite{1989PhRvD..40.3221G}. From Eq.~\eqref{eq:R-adiabatic}, it is evident that the \textit{exact} collapse criterion for the case of vanishing nongravitation attraction is $N/N_{\rm sg} \geq 2/(3\sqrt{3}) \approx 0.3849$, which is slightly lower than the criterion $N/N_{\rm sg}\geq 1$ of Ref.~\cite{1989PhRvD..40.3221G}. In the general case, the runaway collapse criterion is
\begin{align}\label{eq:collapse-criterion}
    \frac{\pd E}{\pd R} \leq 0 \,.
\end{align}
When this condition is satisfied, not only the DM cluster but also the SM enclosed by the DM cluster drop out of local thermodynamic equilibrium.

\subsection{From kinematics to dynamics}\label{sec:kinematics}
Assuming elastic scatterings with nondegenerate SM, the mean variation of the total DM energy per DM-SM scatter is 
\begin{align}\label{eq:Delta-E}
    \Delta E = \frac{1}{(2\:\!\pi)^2}\!\int_0^{2\:\!\pi}\!\!\!\int_0^{2\;\!\pi}\!\!\!\Delta E(\theta,\theta_*) \;\!d\theta\;\!d\theta_* 
\end{align}
where the integrals are taken over the incidental angles $\theta$ and $\theta_*$ with respect to the line of centers and where (see Appendix~\ref{sec:Delta-E} for a derivation)
\begin{align}\label{eq:Delta-E-rel}
    \Delta E(\theta,\theta_*) =&\; 2\;\!\Big[\mathcal{E} p_*^2 \cos^2\!\theta_* - \mathcal{E}_*\;\!p^2 \cos^2\!\theta \nonumber \\
    &\; + (\mathcal{E}\!-\!\mathcal{E}_*)\;\!p\;\!p_* \cos\theta\;\!\cos\theta_* \Big] \nonumber \\
    &\; \times \Big[(\mathcal{E}+\mathcal{E}_*)^2-(p\cos\theta+p_*\cos\theta_*)^2\Big]^{-1}
\end{align}
is the energy gain of DM particles colliding with nuclei for given initial total relativistic energies $\mathcal{E}=K+m$ and $\mathcal{E}_*=K_*+m_*$ (including rest mass but not potential energy), and momenta $p = (\mathcal{E}^2\!-\!m^2)^{1/2}$ and $p_* = ({\mathcal{E}_*}^{\!2}\!-\!{m_*}^{\!2})^{1/2}$.
In the nonrelativistic limit, Eq.~\eqref{eq:Delta-E} reduces to 
\begin{align}\label{eq:Delta-E-non-relativistic}
    \Delta E = - \frac{\beta_+}{2}\big(K\!-\!K_*\big)
\end{align}
where $\beta_+\equiv 4\:\!m\:\!m_*/(m\!+\!m_*)^2$. 
Equation~\eqref{eq:Delta-E-non-relativistic} reduces to the usual formula $\Delta E = \beta_+ K/2$  \citep[e.g.][]{1985ApJ...294..663S},  valid whenever the temperature of the star can be neglected, e.g. during capture and the initial phase of thermalization.
Note that in stars with degenerate nuclear matter (e.g. NSs), considerations are fundamentally different, because postcollision energy states $<E_{\rm F}$ are Fermi blocked for nuclear particles \cite{1989PhRvD..40.3221G}.

The mean timescale between DM-SM scatters is
\begin{align}\label{eq:Delta-t}
    \Delta t  = \frac{1}{n_*\;\! \sigma_*\;\! v_{\rm rel}} \,,
\end{align}
where $n_*=\rho_*/m_*$ is the number density of SM, as seen by a nonrelativistic observer, and $v_{\rm rel}$ is the mean relative velocity. For the present purpose, DM particles can be considered as nonrelativistic. 
The mean relative velocity in Eq.~\eqref{eq:Delta-t} can be shown to be (see, e.g., Ref.~\cite{2017IJMPA..3230002C}, and Appendix~\ref{sec:collision-rate})
\begin{align}\label{eq:vrel-mean}
    v_{\rm rel} = \frac{2\;\![(1+\zeta)^2\;\!K_3(\xi)-(\zeta^2-1)\;\!K_1(\xi)]}{\xi\;\! K_2(x)\;\!K_2(x_*)} \,,
\end{align}
where $\xi \equiv x+x_*$ and $\zeta \equiv (x^2\!+\!x_*^2)/2\:\!x\:\!x_*$ are auxiliary variables and 
$x\equiv m/T$ and 
$x_*\equiv m_*/T_*$ are the standard thermal variables, $K_i(x)$ is the modified (or hyperbolic) Bessel function of the second kind (not to confuse with the kinetic energy that we denote $K$ as well). In the nonrelativistic limit, 
$v_{\rm rel} = [8(mT_*\!+\!m_*T)/\pi\:\! m\:\! m_*]^{1/2}$. 

Assembling the pieces, we obtain the time variation of the total DM mean \textit{per particle} energy from the ratio of Eqs.~\eqref{eq:Delta-E} and \eqref{eq:Delta-t}
\begin{align}\label{eq:dEdt}
    \frac{dE}{dt} = \frac{\Delta E}{\Delta t} \,.
\end{align}
On the other hand, dividing Eq.~\eqref{eq:dE} by $dt$, we obtain the time derivative of the collapse scale
(for $\pd E/\pd R\neq 0$)
\begin{align}\label{eq:dRdt}
    \frac{dR}{dt} = \Big(\frac{\pd E}{\pd R}\Big)^{\!-1}\Big(\frac{dE}{dt} - \frac{\pd E}{\pd N}\frac{dN}{dt}\Big) \,,
\end{align}
where the terms on the right-hand side are given by Eqs.~\eqref{eq:dEdR}, \eqref{eq:dEdt}, \eqref{eq:dEdN} and \eqref{eq:dNdt}, respectively.

\subsection{Heat diffusion}\label{sec:thermodynamics}
Starting from the diffusion equation of stellar specific thermal energy $e_*=K_*/m_*$ (Fick's second law)
\begin{align}\label{eq:Ficks-second-law}
    \frac{\pd e_*}{\pd t} - D \;\!\bs{\nabla}^2 e_* = [\text{heat}\,\text{sources}] \,,
\end{align}
where $D=\kappa/c_p\:\!\rho_*$ is the thermal diffusivity, where $\kappa$ is the thermal conductivity and $c_p$ is the specific heat capacity at constant pressure. On the right-hand side of Eq.~\eqref{eq:Ficks-second-law} we have heat release from DM-SM scattering and, potentially, nuclear reactions' heat release. DM-SM scatterings liberate $N$ times the energy rate $-dE/dt$, given by Eq.~\eqref{eq:dEdt}, per stellar mass $4\pi\rho_* R^3/3$.

Using the finite difference approximation, the Laplacian at $r=0$ can be written as
$\bs{\nabla}^2 e_*|_{r=0}\simeq -6\:\!(e_*\!-\!e_{*\infty})/R^2$ \citep[p.149]{Crank1975TheMO}.\footnote{To convince oneself, one can naturally assume that the specific energy profile is $e_* \propto \exp(-r^2/2R^2)$ for a Gaussian source, then the Laplacian at $r=0$ is $\bs{\nabla}^2e_*|_{r=0}= -e_*/R^2$.}
With this approximation, we obtain a closed form for the time derivative at $r=0$ of the stellar specific energy
\begin{align}\label{eq:desdt}
    \frac{\pd e_*}{\pd t} = -\frac{6\:\!D\:\!(e_*\!-\!e_{*\infty})}{R^2} -\frac{3\:\!N}{4\pi \rho_*R^3}\frac{dE}{dt} + \sum_{i}\dot{q}_{i}\,,
\end{align}
where the terms on the right-hand side account for contributions from (1) diffusion cooling, (2) DM scattering heating remembering that $dE/dt<0$ during runaway collapse, and (3) nuclear reactions' heating, where $\dot{q}_{i}$ is the specific energy generation rate due to nuclear reactions of species $i$. 

The system is closed with the specification of an adequate equation of state $f(\rho_*, T_*,P_*, e_*) = 0$.
Thus, equations \eqref{eq:dNdt}, \eqref{eq:dRdt}, and \eqref{eq:desdt} constitute a closed set of first order differential equations, that determine the evolution of DM collapse in the elastic regime, i.e. the cluster particle number $N(t)$ and its radius $R(t)$, as well as the evolution of specific energy of nondegenerate SM at finite heat diffusion $e_*(t)$.

In this analysis, we have neglected the effect of pressure increase due to heating of stellar matter. This assumption is valid for WD stars where pressure is dominated by electron degeneracy, but might not be valid in main sequence stars, and density modifications can be computed with the help of TOV equation (see, e.g., Ref.~\cite{2016PhRvD..93b3508K} for a study of DM collapse in main sequence stars).

We note that it is also possible to determine the nuclear specific energy at each integration step using Fick's first law (see, e.g., Ref.~\cite{1969ApJS...18..297H}), however, with the cost of solving, at each integration step, a nonlinear find root procedure which can be quite time consuming, specially in the relativistic regime where find root coefficients ($\Delta E$) are numerical integrals.

\subsection{Final stages}

\subsubsection{Fireball evolution}\label{sec:fireball}
The reversible collapse process presented so far is valid as long as scatterings are predominantly elastic. Once the post-collision energy of stellar particles exceeds $200\,$MeV (Hagedorn limit, $T_{\rm H} \sim 1.7\times 10^{12}$K), baryonic particle creation is favored over further heating of stellar matter, and a region of quark-gluon plasma (QGP), also called fireball, is created \citep{1980PhR....61...71S}.  

We expect that collapse then enters an irreversible (``inelastic") stage that quickly leads to the formation of a BH. This assumption is justified unless partons (quarks and gluons) become degenerate and their Fermi sea filled up to mean DM kinetic energy. Since at fireball formation, DM particles are already extremely energetic, we argue that this situation does not occur, though care should be taken. A thorough investigation of the fireball regime exceeds by far the scope of the present article and should be treated in a dedicated study. 

We caution again that DM collapse might follow very different rules in the centers of NSs where matter is expected to be in a degenerate QGP state from the beginning (see, e.g., \cite{2021JCAP...11..056A}).

\subsubsection{BH evolution}\label{sec:swallow}
A BH is formed when $R/N<2Gm$ and $N>N_{\rm Ch}$, where $N_{\rm Ch}\simeq (M_{\rm Pl}/m)^d$ is the Chandrasekhar number where  $d=2$ for bosons~\citep{2012PhRvD..85b3519M} and $d=3$ for fermions~\citep{2011PhRvD..83h3512K}, respectively, and where
$M_{\rm Pl}=1/\sqrt{G}$ is the Planck mass~\citep{2000NuPhB.564..185M}. For all considered DM models (see Sec.~\ref{sec:WD}) and assuming no substructure (see, e.g., Ref.~\citep{2018PhRvL.120x1102S} for a counterexample), we find that the DM cluster does not become degenerate.

The BH's initial mass is $M=Nm$ (neglecting the tiny amount of stellar matter engulfed in the process), and its temporal evolution is given by
\begin{align}\label{eq:dMdt}
    \frac{dM}{dt} = \frac{4\;\!\pi\;\!\rho_*\;\!G^2 M^2}{c_*^3} - \frac{1}{15360\;\!\pi\;\!G^2 M^2} + m\;\!\frac{dN}{dt}\,,
\end{align}
where the terms on the right-hand side account for contributions from (1) Bondi accretion, where $c_*$ is the local sound speed of the star, (2) Hawking radiation, and (3) DM capture and thermalization, where $dN/dt$ is given by equation \eqref{eq:dNdt}. 
If the second term dominates, the BH evaporates ($M\to 0$) and a new DM collapse cycle begins. Otherwise, the star is swallowed by the BH ($M\to M_*$).

We anticipate here noting that in WD stars, for $m \gtrsim 10^{11}$GeV, the mean free path between ions, $(\rho_*/m_*)^{-1/3}$, exceeds the BH's sound horizon, $2 G N m/c_s^2$, and Bondi accretion in Eq.~\eqref{eq:dMdt} is no longer valid. In this regime, accretion is either nearly collisionless or quantum (see, e.g., Ref.~\cite{Giffin:2021kgb}, and references therein). However, we have checked that these effects have little impact on the implosion/evaporation limit in Fig.~\ref{fig:m-sigma-1}.

\section{Phenomenology with WD stars}\label{sec:WD}

\subsection{Input physics}

\subsubsection{Nuclear structure}
We assume WDs composed of equal parts of carbon and oxygen, with mean mass number $A=14$. 
The cross-section between DM and SM particles, $\sigma_*$, depends on the scattering momentum transfer, $\Delta p$, and the de Broglie wavelength of nuclei, $\Delta r = A^{1/3}r_n$, where $r_n=1.25\,$fm is mean separation between nucleons (protons or neutrons) in the nucleus.
If $\Delta p\;\!\Delta r \geq 1/2$, scattering is coherently enhanced, $\sigma_{\!A} \simeq A^2[3 j_1(y)/y]^2 {\rm exp}(-y^2/3)\;\!\sigma_n$, where $\sigma_n$ is the DM-nucleon cross-section, 
$y\equiv 2\;\!\Delta p\;\!\Delta r$ and $j_1$ is the Bessel function of the first kind \citep{1996APh.....6...87L,1988ARNPS..38..751P}. Otherwise, $\sigma_{\! A} = A\;\! \sigma_n$. We neglect additional form factors (see, e.g, Ref.~\cite{2019PhRvD.100d3020A}, for a very detailed discussion).

\subsubsection{Nuclear reactions}
At the time of collapse, WDs have cooled down and crystallization has started from the center \citep{2019A&A...625A..87C}. We model the background core temperature by a simple fitting formula $T_{*\infty} \simeq {\rm min}[10^8,\,3\times 10^6\,(t/{\rm Gyr})^{-1}]\,$K (see, for example, \cite{2011MNRAS.413.2827C}).
The radial temperature profile of the heated WD material drops quickly to the background core temperature $T_{*\infty}$ on a scale of the characteristic radius $R$, such that elemental diffusion is negligible between the heated region and the outside.%
\footnote{It is easy to show that the temperature profile drops with $\propto 1/r$ outside the region where heat is released. Also, if the star is at a temperature $\sim 10^7$, nuclei are stuck in a crystal lattice.} Therefore, the concentration $X_i$ of species $i$ drops with time according to 
\begin{align}\label{eq:dXidt}
    \frac{dX_i}{dt} = - \rho_*\frac{\lambda_{i}}{\bar{M}_i} \,,
\end{align}
where $\lambda_{i}$ and $\bar{M}_i$ are reaction rate and mean molar mass of species $i$, respectively. 

In the present analysis, we limit our investigation to the $^{12} $C$(\gamma,\alpha)^{12}$C reaction, since it is the most interesting for SN Ia phenomenology. 
We use the reaction rate of Ref.~\cite{1988ADNDT..40..283C}, and assume initially $X_{\rm C}=0.5$. The specific nuclear energy generation rate can be written as \cite{1988ADNDT..40..283C}
\begin{align}
    \dot{q}_{i} = f_i\;\! \rho_*\;\! \frac{N_{\rm A}\bar{Q}_{i}}{2}\;\!\frac{X_i^2}{\bar{M}_i^2}\lambda_{i}\,,
\end{align}
where $N_{\rm A}$ is Avogadro's constant, $\bar{Q}_i$ is the mean energy liberated per reaction, and $f_i$ is a factor accounting for electron screening. For carbon fusion, $\bar{Q}_{\rm C} \simeq 3\,$MeV, $\bar{M}_{\rm C}=12\;\!$g$\,$mol$^{-1}$ and $f_{\rm C} \simeq \exp[3.5\;\!(\rho_*/10^9$g$\,$cm$^{-3})^{1/3}(T_*/10^9$K$)^{-1}]$ \cite{1969Ap&SS...5..180A}.

\subsubsection{Heat diffusion}
WDs have thermal diffusion dominated by relativistic electrons when $\rho_*\gtrsim 10^6\;\!$g$\,$cm$^{3}$ \citep{1986bhwd.book.....S}.
We use an interpolation of the results of \citep[][]{Potekhin:2015qsa}\footnote{ \url{http://www.ioffe.ru/astro/conduct/}} for the thermal conductivity, finding that $\kappa \simeq 2.4\times 10^{17} \,$erg$\,$cm$^{-1}$s$^{-1}$K$^{-1}(\rho_*/10^8$g$\,$cm$^{-3})^{1/2}(T_*/10^7$K$)^{1/2}$, strictly valid for $T_* \in  [10^3, 10^9]\,$K and $\rho_* \in [10^{-6}, 10^9]\,$g$\,$cm$^{-3}$. In the absence of predictions for higher temperatures, we extrapolate their results, finding, reassuringly, agreement with $\kappa_{\rm QGP}\simeq 10^{20}\,$erg$\,$cm$^{-1}$s$^{-1}$K$^{-1}$ \citep[e.g.][]{2010PhRvC..81d5205B} at the Hagedorn temperature. The equation of state of ideal gas ions is simply $e_* = 3T_*/2 m_*$ while the specific heat capacity at constant pressure is $c_p = 5/2A$  \citep{1986bhwd.book.....S}.

\subsubsection{Numerical integration}

We integrate the system ($N$, $R$, $e_*$ and $X_{\rm C}$) according to the previously derived equations \eqref{eq:dNdt}, \eqref{eq:dRdt}, \eqref{eq:desdt} and \eqref{eq:dXidt} using a 4th order Runge-Kutta method with adaptive time step. Initial conditions are $(0,R_{\rm th},e_{*\infty},0.5)$.  We investigate three WD masses: $0.6$, $1.0$ and $1.4$ solar masses.

\begin{figure*}
    \centering
    \includegraphics[width=\textwidth]{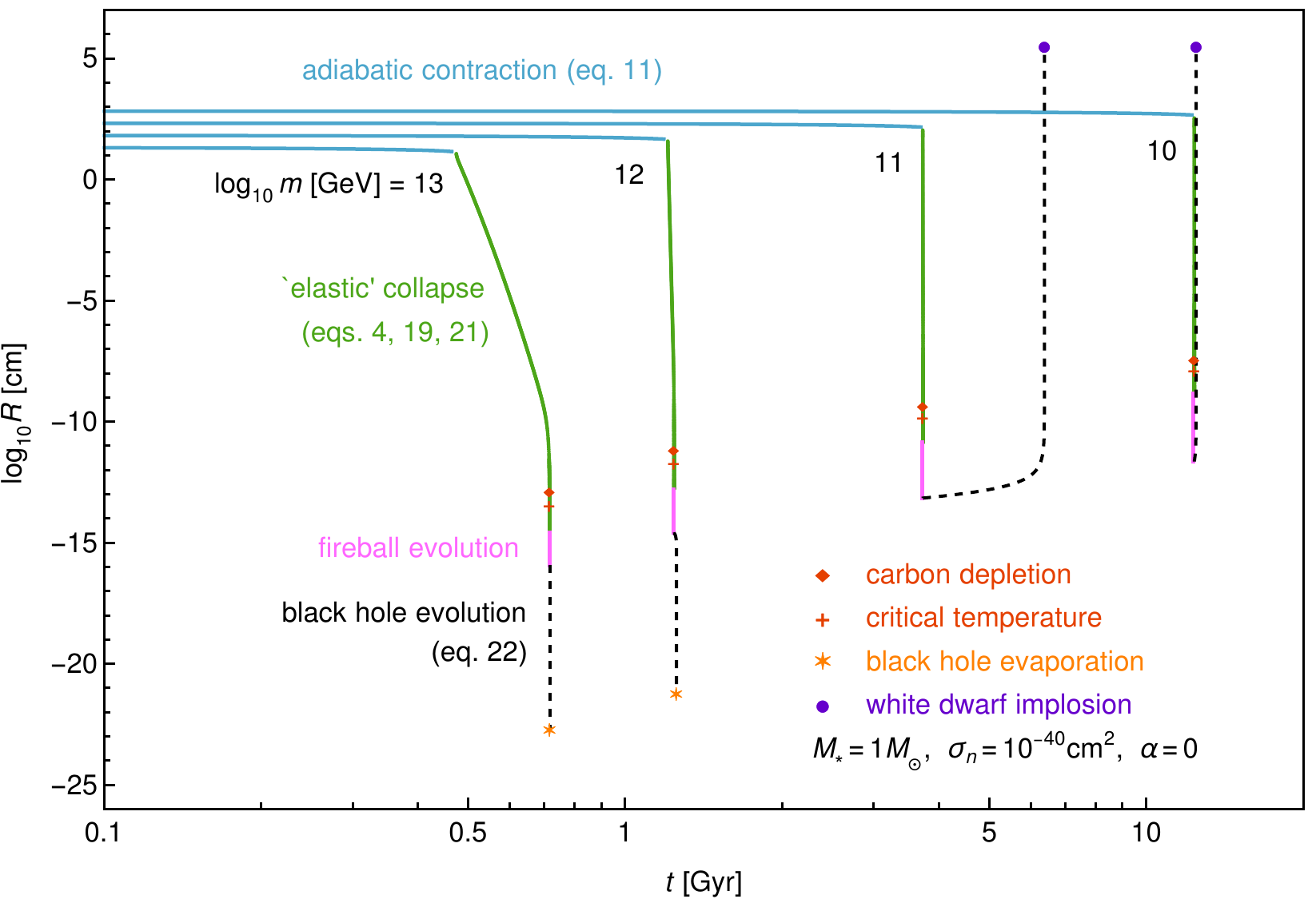}
    \caption{Radius $R$ of the collapsing DM sphere versus age $t$ of the host star, a 1~$M_{\odot}$ WD, for various DM masses $\log_{10}(m/$GeV$)$ as indicated by numbers and DM-nucleon cross section $\sigma_n=10^{-40}\,$cm$^2$. Lines represent adiabatic contraction (cyan), runaway elastic (green) and inelastic (pink) collapse, and BH evolution (black dashed). Carbon depletion and the critical temperature $4.3\times 10^9$K are shown as red diamond and red plus sign, respectively. The destiny of the system is either WD implosion (purple disk) or BH evaporation (yellow star). In the latter case, the process is cyclic (omitted for clarity).
    }
    \label{fig:evolution}
\end{figure*}

\begin{figure}
    \centering
    \includegraphics[width=\linewidth]{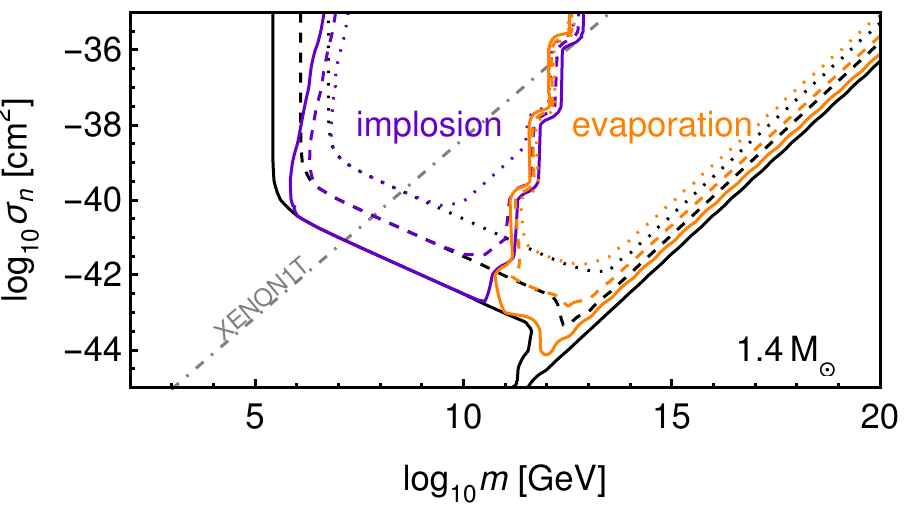}
    \includegraphics[width=\linewidth]{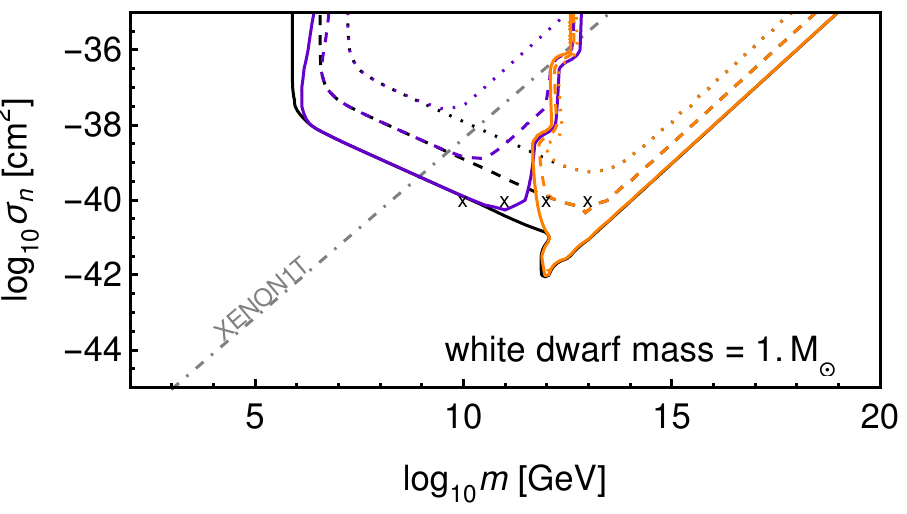}
    \caption{Top panel: parameter regions where DM collapse leads to a mini BH (black) inside a WD with $M_*=1.4~M_{\odot}$, $\alpha=0$, $\mu=0$, and either evaporates (orange) or causes the star to implode (purple) in less than 0.1$\,$Gyr (dotted), 1.0$\,$Gyr (dashed) and 10$\,$Gyr (full).  Including a self-interaction with amplitude $\alpha=10^{-3}$ and range $\mu=1\,$MeV, does not change these parameter regions. The region above the gray dot-dashed line is excluded from XENON1T spin-independent $2\sigma$ bound on DM-nucleon scattering \citep{2018PhRvL.121k1302A}.
    Bottom panel: same as top panel but for $M_*=1.0~M_{\odot}$. The small ``x"s correspond to the evolutionary paths traced in Fig.~\ref{fig:evolution}.
    }
    \label{fig:m-sigma-1}
\end{figure}

In Fig.~\ref{fig:evolution}, we illustrate the time evolution of the collapse scale for some selected DM models highlighting the previously debated collapse phases.
In Fig.~\ref{fig:m-sigma-1} we show the parameter regions where collapse leads to runaway (black), WD implosion (purple) and Hawking evaporation (orange) in less in less than 0.1$\,$Gyr (dotted), 1.0$\,$Gyr (dashed) and 10$\,$Gyr (full) for a 1.4~$M_{\odot}$ WD (top panel) and a 1.0~$M_{\odot}$. Evidently, the most stringent constraints can be obtained with heavy and old WDs.

\subsection{Observational signatures}

\subsubsection{Thermonuclear SN Ia ignition?}\label{sec:thermonuclear}
According to Ref.~\cite{1992ApJ...396..649T}, thermonuclear runaway fusion (deflagration) can proceed if a mass of carbon 
\begin{align}\label{eq:m_C}
    m_{\rm C} \simeq \frac{4}{3}\pi\:\!R^3 \rho_*\;\!X_{\rm C} \in [10^{-5},\,10^{15}]\,{\rm g}
\end{align}
is heated to the critical temperature 
\begin{align}\label{eq:T92}
    \Big(\frac{T_*}{4.3\times 10^9\,{\rm K}}\Big)^{70/3} \gtrsim \Big(\frac{\rho_*}{10^8\,{\rm g}\,{\rm cm}^{-3}}\Big)^{1/2}\Big(\frac{m_{\rm C}}{1\,{\rm g}}\Big)^{-1}\,,
\end{align}
where in the first equality of Eq.~\eqref{eq:m_C} we have assumed that the whole region enclosing the DM cluster is heated to temperature $T_*$. 
Checking the runaway criteria of Eqs.~\eqref{eq:m_C} and \eqref{eq:T92} at each integration step, we find that for all investigated WD masses and DM models (including a Yukawa type self-interacting), stable nuclear burning exhausts the combustible before runaway criteria can be met.  
The explanation resides in the reversible nature of DM collapse based on elastic scatterings, mathematically best appreciable in Eq.~\eqref{eq:Delta-E-non-relativistic}: if $K_*>K$, then $\Delta E>0$, and DM particles gain energy. In practice, from our numerical simulations, we find that once nuclei become almost as hot as DM particles, the collapse process slows down or interrupts momentarily until exothermic nuclear reactions are over.

\subsubsection{Pycnonuclear SN Ia ignition?}\label{sec:pycnonuclear}
Pycnonuclear (density driven) carbon reactions start if the nuclear density exceeds $\rho_* \gtrsim 3\times 10^9\,$g$\,$cm$^{-3}$. 
Solving the TOV equation with the addition of a top hat DM density profile and assuming zero temperature equation of state, we find that sizable density increase never occurs before thermonuclear reactions.
However, after the passage of the thermonuclear flame, the central region of the WD is carbon depleted, while the outer shells of the core are crystallized and elemental diffusion suppressed. Since density increase at later stages encloses a smaller region than the carbon depleted one, pycnonuclear ignition remains illusive. This is a conservative estimate, since assuming the hot equation of state, additional ideal gas pressure counters the gravitational pull, and it is questionable if density is increased at all.

\subsubsection{Constraints from existing WDs}
With current telescopes, WDs can only be observed directly in the neighborhood of the Sun. These experience a DM density $\rho_{\rm gal,0}\simeq 0.4\,$GeV/cm$^{-3}$ and a velocity dispersion $v_{\rm gal,0}\simeq 200\,$km$\,$s$^{-1}$. The best constraints come from heavy and old observed WDs (see Table~\ref{tab:wd}).

\begin{table}
    \centering
    \begin{tabular}{lccr}
    \hline
    Name $\!\!$    &  mass  &  cooling age & reference 
    \\
    \hline 
    WD 0346 $\!\!$  & $0.77\,M_{\odot}$ &  $11.0\,$Gyr & \protect\cite{2012MNRAS.423L.132K} 
    \\
    WD 1832+089 $\!\!$ & $1.33\,M_{\odot}$ & $330\,$Myr & \protect\cite{2020MNRAS.499L..21P} 
    \\
    SDSS J2322+2528 & $1.13\,M_{\odot}$ & $4.58\,$Gyr & \!\!\!\protect\cite{2016MNRAS.455.3413K,2018yCat.1345....0G,2019AA...625A..87C} 
    \\
    \hline
    \end{tabular}
    \caption{Solar neighborhood WDs that constrain asymmetric DM models, see Fig.~\ref{fig:constraints} (only the two most constraining are shown). 
    }
    \label{tab:wd}
\end{table}

\begin{figure}
    \centering
    \includegraphics[width=\columnwidth]{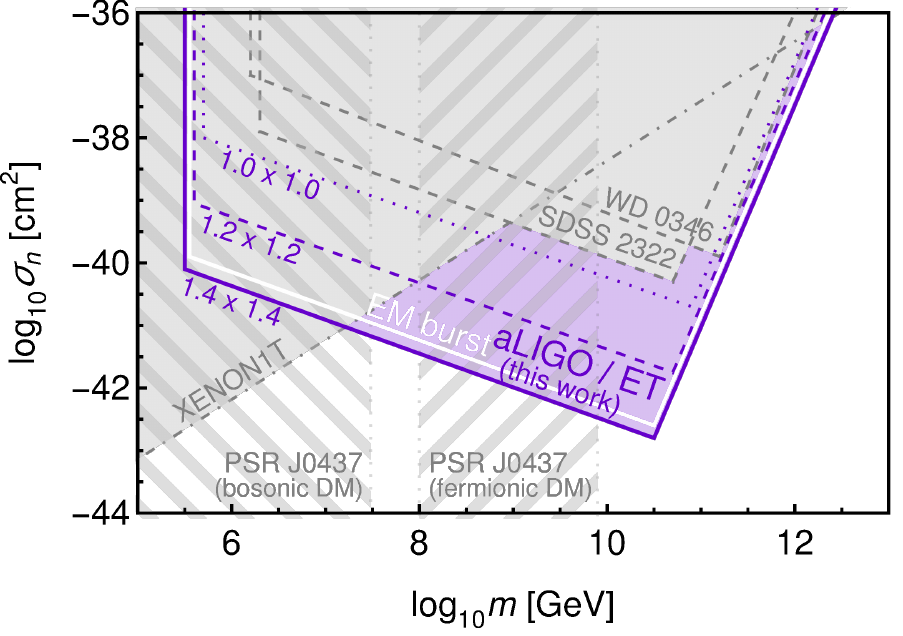}
    \caption{Constraints on asymmetric DM. The gray shaded regions are excluded from nondetections with the XENON1T experiment (dot-dashed contour; \citep{2018PhRvL.121k1302A}), from the existence of various old WDs in the solar neighborhood (dashed contour, see Table~\ref{tab:wd} for details). The diagonal (antidiagonal) hatched region is excluded from the existence of pulsar PSR J0437-4715 in case of bosonic (fermionic) DM \citep{2021PhRvL.126n1105D}.
    The purple shaded region is marginally excluded from the nondetection of binary BHs by aLIGO/Virgo, originating from DM collapse-induced implosion of binary WDs with component masses as indicated by numbers and $\sim 10$~Gyr merger delay times; future facilities like the Einstein telescope (ET) can confirm these constraints, see Table~\ref{tab:GW}. The parameter region above the white full line is excluded due to the nonobservation of specific electromagnetic (EM) bursts following the implosion of magnetic WDs (see Sec.~\ref{sec:EM-bursts} for details).
    Mean Galactic DM parameters  ($\rho_{\rm gal}\sim 0.4\,$GeV/cm$^3$ and $v_{\rm gal}\sim 200\,$km/s) have been assumed.}
    \label{fig:constraints}
\end{figure}

Interpolating the previously obtained results, we show in Fig.~\ref{fig:constraints} the DM models that are excluded simply because these WDs did not implode to a BH. As can be seen, these constraints are competitive with current direct detection experiments in the mass range $10^9-10^{12}\,$GeV where they roughly exclude DM models with DM-nucleon cross section $\sigma_n \gtrsim 10^{-40}\,$cm$^2$.

SDSS J2322+25328 is the most constraining. Heavier and, at the same time, older have not been found so far, but are expected to exist, especially in the thick disc and stellar halo components of the Galaxy which were formed before approximately $11\,$Gyr. 
It is possible that these have imploded under the pull of DM in the green parameter region (Fig.~\ref{fig:constraints}).

\subsubsection{Electromagnetic bursts}\label{sec:EM-bursts}

The recent 20~pc volume-limited survey has revealed a high incidence (22\%) of magnetic WDs \citep{2021MNRAS.507.5902B}. Most (85\%) of these concern single WDs, and the field strength distribution is logarithmically uniform in the range $\sim 4\times 10^4$G--$10^9$G \citep{2021MNRAS.507.5902B}. Besides, no evidence for a correlation between field strength and WD mass, neither any sign for field strength decay with time have been found \citep{2021MNRAS.507.5902B}.   

According to the no-hair theorem, which prevents magnetic field lines from puncturing the event horizon, the newly formed BH must expel its magnetic field \citep{2013PhRvD..88d4020D}, liberating the energy contained in the magnetosphere, at least
\citep[e.g.][]{2014A&A...562A.137F}
\begin{align}\label{eq:E_B}
    E_B \sim &\; \frac{B^2}{8\pi}\frac{4\pi}{3}R_*^3 \nonumber \\
    \simeq &\; 6\times 10^{44}~{\rm erg}~ \Big(\frac{B}{10^8{\rm G}}\Big)^2\Big(\frac{R_*}{4\times 10^8{\rm cm}}\Big)^3\,,
\end{align}
where $B$ is the surface magnetic dipole field strength of the WD and  $R_*$ its radius (in Eq.~\eqref{eq:E_B}, the radius of a  1.2-$M_{\odot}$ WD is shown). The relevant timescale of the final stage of implosion is the free-fall time \citep{2003ApJ...585..930B,2014A&A...562A.137F}
\begin{align}\label{eq:tff}
    \Delta t_{\rm ff} \sim &\; \Big(\frac{R_*^3}{8 G M_*}\Big)^{1/2} \nonumber \\
    \simeq &\; 225~{\rm ms}~\Big(\frac{M_*}{1.2~M_{\odot}}\Big)^{-1/2}\Big(\frac{R_*}{4\times 10^8~{\rm cm}}\Big)^{3/2}\,,
\end{align}
and we assume that this sets the duration of the energy emission. 
Magnetohydrodynamics simulations of nonrotating magnetic NSs show that about 5\% of the available energy is emitted in the main burst \citep{2013PhRvD..88d4020D}. We assume that a similar fraction is emitted in the case of WDs, leading to a luminosity of
\begin{align}
    L_B \sim &\; \eta \;\! \frac{E_B}{\Delta t_{\rm ff}} \simeq  1.3\times 10^{44}~{\rm erg}~{\rm s}^{-1}\Big(\frac{\eta}{0.05}\Big)\Big(\frac{B}{10^8{\rm G}}\Big)^2\nonumber \\
    &\; \times \Big(\frac{M_*}{1.2~M_{\odot}}\Big)^{1/2}\Big(\frac{R_*}{4\times 10^8{\rm cm}}\Big)^{3/2}\,,
\end{align}
where $\eta$ is an efficiency factor.

According to Eq.~\eqref{eq:E_B}, the most energetic bursts disrupt in WDs with the strongest magnetic fields.
Assuming $10^{10}$ WDs in the Galaxy \citep{2009JPhCS.172a2004N}, we estimate that $6\times 10^6$ single WDs have masses heavier than 1.2~$M_{\odot}$ and magnetic dipole fields stronger than $10^8$~G \citep{2020ApJ...898...84K,2021MNRAS.507.5902B}. 
Adopting the Galactic stellar structure model of Ref.~\cite{2003A&A...409..523R}, the fraction of WDs exposed to a DM density larger than the local DM density and taking into account only thick disc and stellar halo WDs (which are $10\,$Gyr old) is 52\%. Assuming that the initial star burst lasted $\sim 1$~Gyr \citep{2021MNRAS.502.1753T}, and considering that the number density of MW-like galaxies is $10^{-2}$Mpc$^{-3}$, we find a volumetric burst rate of $1.3\times 10^{-4}$~Mpc$^{-3}$~yr$^{-3}$. 

The details of how this energy is converted into radiation is uncertain (see, for example, Refs.~\citep{2012PhRvD..86j4035L,2014A&A...562A.137F} for options ranging from gamma-ray to radio bursts). 
For instance, the bulk of fast radio bursts (FRBs) emit a luminosity in the range $10^{41}$--$10^{44}$~erg~s$^{-1}$ \citep[e.g.][]{2022ApJ...926..206Z} and their volumetric rate is $10^{-4}$~Mpc~yr$^{-3}$ \citep{2013Sci...341...53T}. Thus, the rate of the most energetic ($>10^8$G) bursts from WD implosions alone equals the total rate of all known FRBs taken together. Clearly, these WD implosions should have been noticed. 
When it comes to durations, typical FRBs last 1--10~ms with none longer than 100~ms detected so far, and nonrepeating are typically shorter than repeating \citep{2019Natur.566..230C}. According to Eq.~\eqref{eq:tff}, WD implosions expedite longer (50--1000~ms) lasting bursts, roughly 2 orders of magnitude longer. If their emission occurs in radio wave lengths, then the nondetection of long FRBs stringently constrains asymmetric DM models (see Fig.~\ref{fig:constraints}). 

Gamma ray bursts (GRBs) have much wider spread durations ranging from 10~ms to several hours. The particular class of short ($<2$~s) GRBs accounts for 30\% of the total rate, and is associated with regions of little or no star formation, such as large elliptical galaxies and the central regions of galaxy clusters \citep{2006ApJ...638..354B}. This rules out a link to massive stars, but makes them eligible for emission from the transmutation of old magnetic WDs. 
The commonly accepted mechanism of short GRBs is the merger of two NS \citep{2007PhR...442..166N} or the merger of a NS with a BH, which is consistent with minutes to hours lasting afterglows, caused by fragments of tidally disrupted material remaining in orbit while inspiraling into the BH over a longer period of time. On the other hand, these afterglows are difficult to explain with WD implosions. 

In sum, neither FRBs nor GRBs match with the expected properties of DM collapse induced WD implosions. We note that the field energy estimation in Eq.~\eqref{eq:E_B} is an absolute lower bound, because in ideal magnetohydrodynamics, the field is ``frozen-in" with the fluid and increases linearly with density (while Eq.~\eqref{eq:E_B} assumes constant field strength); analytical and numerical calculations in Newtonian gravity and general relativity show that internal magnetic field strengths of up to $10^{12-16}$~G are possible (see  \citep{2018MNRAS.477.2705B} and references therein). However, these simple energy arguments should be tested in magnetohydrodynamics simulations of imploding WDs.

\subsubsection{Detection of GWs from sub-Chandrasekhar BHs}

The GW signature of binary WD mergers is very different from that of binary BH mergers with the same mass. Typically, the secondary (lighter and larger) is spaguetified by tidal forces of the primary (heavier and more compact) prior to coalescence. The orbital motion is expected to be observed at sub-Hz frequencies by future space-born laser interferometric detectors of gravitational waves \citep{2021ApJ...906...29Y}, while super-Hz emission of binary BH mergers is already detectable by aLIGO \citep{2018PhRvL.121w1103A}.

In Table~\ref{tab:GW}, we compare the detection rates per year of current and future GW detectors for different combinations of component masses. In order to obtain strong constraints (Fig.~\ref{fig:constraints}), both components must be heavy and have long ($\sim 10$~Gyr) formation-to-merger delays to allow for DM accumulation and subsequent transmutation of both WDs to BHs. Since heavier WDs are also rarer we consider two mass bins, $[1.0,1.2]~M_{\odot}$ and $[1.2,1.4]~M_{\odot}$ (Table~\ref{tab:GW}).

\begin{table}
    \centering
    \begin{tabular}{llrrr}
    \hline 
    $M_1$ & $M_2$ & aLIGO & aLIGO-DS & ET \\
    $[M_{\odot}]$ & $[M_{\odot}]$ & [yr$^{-1}$] & [yr$^{-1}$] & [yr$^{-1}$] \\
    \hline
    $[1.0,1.2]$ & $[1.0,1.2]$ & 0.6 & 16 & 3~025 \\
    $[1.2,1.4]$ & $[1.0,1.2]$ & 0.3 & 7 & 1~323 \\
    $[1.2,1.4]$ & $[1.2,1.4]$ & 0.1 & 3 & 579  \\
    \hline
    $[1.0,1.4]$ & $[1.0,1.4]$ & 1.2 & 33 & 6~250 \\
    \hline
    \end{tabular}
    \caption{GW detection rates per year (columns 3 to 5) from transmuted binary BH mergers with component masses comprised in the specified bins (columns 1 and 2). The last line is the total rate. These rates assume the following detector ranges for $\mathcal{O}(1)~M_{\odot}$ binary BH mergers: 110~Mpc for aLIGO \protect\citep{2018PhRvL.121w1103A}, 
    330~Mpc for aLIGO at design sensitivity (DS; \protect\citep{2018LRR....21....3A}), and 1.9~Gpc for ET \protect\citep{2022Entrp..24..262N}.}
    \label{tab:GW}
\end{table}

We estimate the rates as follows. The total number of WDs in a MW-like galaxy is $10^{10}$ \citep{2009JPhCS.172a2004N}, and the Galactic merger rate per WD is $10^{-11}$yr$^{-1}$ \cite{2018MNRAS.476.2584M}.
Assuming the 100~pc volume-limited mass function in the SDSS footprint \cite{2020ApJ...898...84K}, the fraction of WDs with mass greater than 1.0~$M_{\odot}$ is 4.4\%; since both binary companions must satisfy this, we have a fraction $(0.044)^2$. Since 30--50\% of high-mass WDs have a merger history \citep{2020A&A...636A..31T}, we have to multiply by an additional factor of $\sim (1-0.4)^2$ (neglecting systems of higher multiplicity than 2).  Results an expected binary WD merger rate of
$7\times 10^{-5}$~yr$^{-1}$ for component masses in $[1.0,1.4]~M_{\odot}$ per MW-like galaxy.

This rough estimate is consistent with detailed binary population synthesis calculations of Ref.~\cite{2010Natur.463...61P}, who compute the rate of binary mergers with primary mass between 0.85 and 1.05~$M_{\odot}$ and a mass ratio $0.9\leq M_2/M_1 \leq 1.0$ to be 2--11\% of the type Ia supernova rate (see supplementary information of \citep{2010Natur.463...61P}). For the type Ia supernova rate from the LOSS survey, $(5.4\pm 0.1)\times 10^{-3}$ \cite{2011MNRAS.412.1473L}, the resulting double WD merger rate is $(3.5\pm 2.4)\times 10^{-4}$~yr$^{-1}$, while our crude estimate for this mass range yields $7\times 10^{-4}$~yr$^{-1}$. The remaining discrepancy of a factor of 2 could be due to the adopted mass function; if we assume the binary population synthesis mass function of Ref.~\cite{2020A&A...636A..31T}, we find a rate $3\times 10^{-4}$~yr$^{-1}$.

Adopting the Galactic stellar structure model of Ref.~\cite{2003A&A...409..523R}, we estimate the fraction of WDs exposed to a DM density larger than the local DM density and taking only into account thick disc and stellar halo WDs (which are $10\,$Gyr old), to 52\%.
Since the main channel for high-mass double degenerates is a single common envelope phase, we estimate that 60\% of these have long formation-to-merger delay times ($\sim 10$~Gyr, see Appendix~\ref{sec:details-merger-rate} for details).
Taking into these cuts, and considering that the number density of MW-like galaxies is $10^{-2}$Mpc$^{-3}$, we finally find the detection rates specified in Table~\ref{tab:GW}.
The nonobservation of sub-Chandrasekhar binary BH mergers over a period of time $\delta t$ would bound their rate to $\leq 2.3/\delta t$ at 90\% confidence level.

The constraints on DM models from capture in binary systems are slightly more stringent than those from single WDs, due to enhanced DM capture rates in binary systems \cite{2012PhRvL.109f1301B}. A maximum enhancement factor of $\sim 4.3$ has been found for orbital periods of 8~h \cite{2012PhRvL.109f1301B}, attributed to the energy loss by DM particles resulting from their gravitational scattering off moving companions. 
Binary WDs that merge in a Hubble time due to gravitational radiation have initial orbital periods of at most 13.5~h \citep{1962ApJ...136..312K}.
Based on the results in Table~1 of Ref.~\cite{2012PhRvL.109f1301B}, and since the binary spends most of its evolution at large orbital periods, we estimate that the integrated amplification factor is $\sim $3--4. This means 3--4 times earlier collapse and the parameter region of constraints from binary systems is correspondingly larger than it would be from single stellar systems (see Fig.~\ref{fig:constraints}).

\subsubsection{Disentangling transmuted from primordial BH inspirals}
As the two BHs merge, the morphology of the resulting gravitational waveform depends on the phenomenological effective inspiral spin parameter \citep{2001PhRvD..64l4013D}
\begin{align}\label{eq:chi-eff}
    \chi_{\rm eff} \equiv \frac{M_1\;\!\chi_1\cos\theta_1 + M_2\;\!\chi_2\cos\theta_2}{M_1+M_2}\,,
\end{align}
where $\theta_1$ and $\theta_2$ are the misalignment angles between the component spins and the orbital angular momentum, $M_1$ and $M_2$ are the component masses, and $\chi_1$ and $\chi_2$ are the dimensionless component spins, defined by $\chi \equiv J/M^2$ and limited to values $\in [0,1]$, where $J=2\pi I/P$ is the angular momentum, $I$ the moment of inertia and $P$ the rotation period. 
Since the angular momentum $J$ is conserved during WD transmutation to a BH, we can estimate $\chi$ from typical values of the progenitor WD star. 

The primary mechanism for producing tight binary systems is a common-envelope evolution \citep{2013A&A...557A..87T,2018MNRAS.481.1908K}, where each component results from single stellar evolution. The rotation periods of typical low mass WDs originating from single stellar evolution are of the order of 1~d, as inferred from rotational broadening of spectral lines \citep{2005A&A...444..565B} and asteroseismology \citep{2015ASPC..493...65K,2017ApJS..232...23H}. For high mass WDs, the rotation periods tend to be much shorter \citep{2017ApJS..232...23H}, consistent with the tendency for faster rotating stellar progenitors to produce heavier cores \citep{2019ApJ...871L..18C}. Lacking number statistics for heavy WDs, we assume here a representative value for the rotation period of 1~h (roughly extrapolating data in the right panel of fig.~8 of Ref.~\citep{2017ApJS..232...23H}), yielding dimensionless spin values
\begin{align}\label{eq:dimensionless-spin}
    \chi \simeq 0.2 \,\Big(\frac{P}{1~{\rm h}}\Big)^{\!-1}\Big(\frac{M_*}{1~M_{\odot}}\Big)^{\!-2}\,,
\end{align}
where we have assumed conservatively the moment of inertia of nonrotating WDs, $I=10^{50}$g~cm$^2$, which has little dependence on mass \citep{2017MNRAS.464.4349B}. 

Binary WDs formed via ``isolated" evolution (the dominant formation channel) have spin vectors which are likely to be closely aligned with the orbital angular momentum, hence $\theta_1 \simeq \theta_2 \simeq 0$. We assume 
$\chi_1 \simeq \chi_2$ and $M_1\simeq M_2$ for simplicity. Then, Eq.~\eqref{eq:chi-eff} reduces to  $\chi_{\rm eff} \simeq \chi$. Using Eq.~\eqref{eq:dimensionless-spin}, we find a representative value $\chi_{\rm eff} \simeq 0.2$, a value that is already measurable with current 90\% credible intervals, $\simeq 0.15$, for the default model \citep[e.g.][]{2021ApJ...921L..15G}. 

In particular, transmuted binary BH inspirals can be disentangled from primordial binary BH binary inspirals. In standard cosmology, the QCD phase transition is expected during a radiation-dominated cosmological epoch and hence these primordial BHs are expected to have very low intrinsic spin magnitude, roughly distributed as a Gaussian peaked at $\chi_{\rm eff}=0$ and with variance $\sigma_{\chi_{\rm eff}} \simeq 0.35$ \citep{2017PTEP.2017h3E01C}. 
We note that while current sensitivity is sufficient to discriminate between these peak values, a certain number of events must be observed to overcome the intrinsic dispersion of $\chi_{\rm eff}$ to disentangle between the different solar mass BH production channels. This further motivates upcoming GW detectors like the Einstein telescope.


\section{Summary and discussion}\label{sec:conclusion}
Runaway collapse of a DM cluster at the center of a star at finite temperature is governed by a system of differential equations. 
We have derived these equations for the relatively simple situation where nuclei are nondegenerate, pressure feedback is small, and DM-nuclear collisions are elastic. 
Nevertheless, these ``elastic" collapse equations, i.e. Eq.~\eqref{eq:dNdt}, Eq.~\eqref{eq:dRdt} and Eq.~\eqref{eq:desdt},  are valid for most of the collapse evolution in WD stars (see Fig.~\ref{fig:evolution}). 
In the presence of nuclear reactions, the system can be coupled with a set of equations governing elemental concentrations, i.e. Eq.~\eqref{eq:dXidt}.

Local heating of nuclear matter from scattering with DM is controlled by finite heat diffusion. When carbon reactions dominate the heat release, further collapse is interrupted until reactions are over. Consequently, when the critical temperature for thermonuclear runaway is reached, carbon is already depleted (Fig.~\ref{fig:evolution}). Thus, in crystallized WDs where elemental diffusion is suppressed, type Ia supernova ignition from DM collapse remains illusive.
Subsequent ignition mechanisms, i.e. when the DM cluster has collapsed to a smaller radius, face a situation where heat release occurs in a region much smaller than the previously carbon depleted.

Instead, a mini BH is formed at the center of the star, and the stellar matter is accreted leaving behind a macroscopic BH. 
In case of total accretion of the star onto the mini BH, several observational signals are detectable with current technology.
First, the mere existence of old and heavy WDs in the solar neighborhood imposes weak but solid constraints on asymmetric DM (see Fig.~\ref{fig:constraints}). 
Second, the nondetection of 50--1000~ms lasting electromagnetic bursts from the ejection of the WD magnetic field upon transmutation to a BH places stringent constraints (see Fig.~\ref{fig:constraints}); these restrictions are uncertain as the exact details of the burst are model dependent (with possibilities ranging from FRBs to GRBs).

Third, the most stringent and solid constraints result from the nondetection of GW signals from binary BH coalescences with sub-Chandrasekhar component mass. We find that
aLIGO/Virgo should detect $\sim 1$ event per year ($\sim 30$ per year at design sensitivity), while future gravitational wave facilities like the Einstein telescope~\citep{2022Entrp..24..262N} would detect $\sim 6~000$ per year. Their exclusion limits are competitive with current direct detection experiments \citep{2018PhRvL.121k1302A} and pulsar constraints \cite{2021PhRvL.126n1105D} in case of bosonic DM; in case of fermionic DM, pulsar constraints are currently more stringent for part of the parameter space (see Fig.~\ref{fig:constraints}).

GWs emitted by transmuted BH mergers can be disentangled from those of primordial BHs with the same mass due to different effective inspiral spin parameter distributions. We have predicted a peak value of $\chi_{\rm eff} \simeq 0.2$ for transmuted origin, while for primordial origin $\chi_{\rm eff} \simeq 0$ is expected. We note that our prediction depends on the relatively uncertain rotation periods of solar mass WDs that originate from single stellar evolution. Ongoing space-based short-cadence photometric missions like TESS and CHEOPS will greatly improve asteroseismic studies \cite{2018A&A...620A.203M}.


\begin{acknowledgements}
We thank the anonymous referee for useful comments and suggestions that led to significant improvements, mainly concerning the section on detections of GWs from binary BHs coalescences. We also thank Davi Rodrigues for useful comments on the manuscript.
H.S. is thankful for FAPES/CAPES DCR grant No. 009/2014. V.M. thanks CNPq and FAPES for partial financial support. S.P. is partly supported by the U.S. Department of Energy, grant No. de-sc0010107.
\end{acknowledgements}




\bibliography{biblio.bib} 



\appendix

\section{POTENTIAL ENERGY}\label{sec:potential-energy}

Consider a system composed of $N$ particles with mass $m$ arranged in a spherically symmetric mass distribution $\rho(r)$. The total potential energy due to its own gravity is
\begin{align}
    U = - 4\;\!\pi\;\! G\!\!\int_{r=0}^{\infty}M(r)\;\!\rho(r)\;\!r\;\!dr\,,
\end{align}
where $M(r)$ is the integrated mass (mass inside $r$) is
\begin{align}\label{eq:M-of-r}
    M(r) = 4\;\!\pi\!\!\int_{r'=0}^{r} \rho(r')\;\!r'^2\;\!dr'\,.
\end{align}
For a top hat distribution ($\rho(r)=\rho$ for $r\leq R$ and $\rho(r)=0$ for $r>R$), we have
\begin{align}
    U = - \frac{3\;\!G\;\!M^2}{5\;\!R}\,.
\end{align}
For a Maxwell-Boltzmann mass distribution, with
\begin{align}
    \rho(r) = \rho_0\;\!\exp\Big(-\frac{r^2}{2\;\!R^2}\Big) = \sqrt{\frac{2}{\pi}}\frac{M}{4\;\!\pi\;\!R^3} \exp\Big(-\frac{r^2}{2\;\!R^2}\Big)\,,
\end{align}
the total potential energy due to its own gravity is
\begin{align}
    U = - \frac{G M^2}{2\;\!\sqrt{\pi}\;\!R}\,.
\end{align}

Now consider the gravitational potential of the system due to an external mass distribution $\rho_*(r)$,
\begin{align}
    U_* = -4\;\!\pi\;\!G\int_{r=0}^{\infty}M_*(r)\;\!\rho(r)\;\!r\;\!dr\,,
\end{align}
where $M_*(r)$ is given by Eq.~\eqref{eq:M-of-r} with stars added to $M$ and $\rho$. 
If the external mass density is constant $\rho_*(r)=\rho_*$ (which is a good approximation for the center of a star), and the system has a top-hat distribution (unrealistic), we have
\begin{align}
    U_*= - \frac{4\;\!\pi\;\!G\;\!\rho_*\;\!M\;\!R^2}{5}\,.
\end{align}
If the system has a Maxwell-Boltzmann density distribution (and $\rho_*(r)$ still constant), we have
\begin{align}
    U_* = - 4\;\!\pi\;\!G \rho_*\;\!M\;\!R^2\,.
\end{align}

The mean  potential energy \textit{per particle} of a system composed by $N$ particles with mass $m$ and a top-hat mass distribution $\rho$ in an external constant distribution $\rho_*$ is (dividing by $N$ and replacing $M=N\;\!m$)
\begin{align}
    \frac{1}{N}\big(U+U_*\big) =  - \frac{3\;\!G N m^2}{5\;\!R} - \frac{4\;\!\pi\;\!G\rho_*\;\!m R^2}{5} \,.
\end{align}
The mean potential energy  \textit{per particle}  of a Maxwell-Boltzmann mass distribution in an external constant distribution $\rho_*$ is
\begin{align}
    \frac{1}{N}\big(U+U_*\big) = - \frac{G N m^2}{2\;\!\sqrt{\pi}\;\!R} - 4\;\!\pi\;\!G\rho_*\;\!m R^2\,.
\end{align}
In the main article we use the letter $U$ for the total mean potential energy \textit{per particle}.

\section{THERMALIZATION TIMESCALE}\label{sec:thermalization}
In this Section, we derive the formula \eqref{eq:t-th}. A derivation of part of the formula has been given previously by \citep{2011PhRvD..83h3512K}. The total thermalization time can be divided into three stages  (1) orbital decrease crossing the star twice every orbital period, (2) orbital decrease completely inside the star with $v>v_*$, and (3) orbital decrease completely inside the WD with  $v<v_*$.

\subsection{First stage}
During the first stage, the DM particle has a chance to loose kinetic energy twice each orbital period $P=2\;\! \pi\;\! \sqrt{a^3/(GM)}$, where $a$ is the semimajor axis and $M$ the mass of the star, such that the timescale between collisions is
\begin{align}\label{eq:Dt-1}
    \langle \Delta t\rangle = \frac{1}{2}\;\! P\;\! \frac{\sigma_{\rm sat}}{\sigma_{\chi A}} = \pi \;\! \Big(\frac{a^3}{G M}\Big)^{1/2}\;\! \frac{\sigma_{\rm sat}}{\sigma_{\chi A}}
\end{align}
where $\sigma_{\rm sat} = R^2\;\! m/M$ is the saturation cross section. 
The total energy of the DM particle with semimajor axis $a$ is
\begin{align}\label{eq:E-th1}
    E = - \frac{G M\;\! m}{a}\,.
\end{align}
Assuming radial orbits and constant density star, the potential energy at a radial position $r$ inside the star is 
\begin{align}
    U =&\; - \frac{G M\;\! m}{R} \Big(\frac{3}{2}-\frac{r^2}{2\;\!R^2}\Big) \,, & (r<R)
\end{align}
where $R$ is the radius of the star. The (instantaneous) kinetic energy is
\begin{align}
    K =&\; E - U = \frac{G M m}{R}\Big(\frac{3}{2} - \frac{r^2}{2\;\!R^2} - \frac{R}{a}\Big)\,, &  (r<R)
\end{align}
Averaging over radial positions, the mean kinetic energy is 
\begin{align}
    \langle K \rangle = \frac{1}{R} \int_0^R K\;\!dr = \frac{G M m}{R}\Big(\frac{4}{3}- \frac{R}{a}\Big)\,.
\end{align}
The mean variation per scatter is (assuming the nonrelativistic limit, Eq.~\eqref{eq:Delta-E-non-relativistic}, and neglecting $K_*$)
\begin{align}\label{eq:DE-1}
    \langle \Delta E \rangle = -  \frac{\beta_+}{2} \frac{G M m}{R}\Big(\frac{4}{3}- \frac{R}{a}\Big)
\end{align}
and, treating scattering as a continuous process,  combining Eq.~\eqref{eq:Dt-1} and Eq.~\eqref{eq:DE-1}, we have
\begin{align}
\frac{dE}{dt} = \frac{\langle \Delta E\rangle}{\langle \Delta t\rangle}
\end{align}
and, from \eqref{eq:E-th1} we have
\begin{align}
    \frac{da}{dE} = \frac{G M m}{E^2} = \frac{a^2}{G M m}
\end{align}
Assembling these equations
\begin{align}
    \frac{da}{dt} = \frac{da}{dE}\;\!\frac{dE}{dt} = - \frac{A_1}{\sqrt{a}}\;\!\Big(B_1\;\!a-1\Big)
\end{align}
where
\begin{align}
    A_1 =&\; \frac{\beta_+ \sqrt{G M}\;\!\sigma_A}{2\pi\;\!\sigma_{\rm sat}}\,,\\
    B_1 =&\; \frac{4}{3R}
\end{align}
Integrating, we have
\begin{align}
    t_1 = \int dt = \frac{1}{A_1}\int_{a_0}^{R} \frac{\sqrt{a}\;\!da}{(B_1\;\!a-1)}
\end{align}
Using $a\;\!B_1 = \cosh^2 x$, we obtain
\begin{align}
    t_1 = \frac{2}{A_1\;\!B_1^{3/2}}\Bigg\lbrace\sqrt{B_1\;\!a_0} - \sqrt{B_1\;\! R} + \frac{1}{2} \ln\Big[\frac{R\;\!(B_1\;\!a_0-1)}{a_0\;\!(B_1\;\!R-1)}\Big]  \Bigg\rbrace
\end{align}
neglecting the logarithm which is $\ln 2$, and assuming $m \ll m_*$, we have $\beta_+\simeq 4m_*/m$, and
\begin{align}
    t_1 \simeq \frac{3\;\!\pi\;\!R\;\!\sigma_{\rm sat}\;\! m}{4\;\!\sigma_A\;\!m_*}\Big(\frac{a_0}{G M}\Big)^{1/2}
\end{align}
Assuming $a_0 = R\;\!v_e^2/v_{\rm gal}^2$, and using $\sigma_{\rm sat} = \pi\;\!R^2 m_*/M \simeq 3\;\!m_*/(4\;\!R\;\!\rho_*)$, we have
\begin{align}
    t_1 \simeq \frac{9\;\!\sqrt{2}\pi\;\! m}{16\;\!\rho_*\;\! \sigma_A\;\!v_{\rm gal}}
\end{align}

\subsection{Second stage}
For orbits totally inside the star ($a<R$), and assuming constant density, the total energy for a DM particle with semimajor axis $a$ is 
\begin{align}\label{eq:E-th2}
    E = &\; -\frac{G M m}{R}\Big(\frac{3}{2}- \frac{a^2}{2\;\!R^2}\Big)\, & (a\leq R)\,.
\end{align}
The (instantaneous) potential energy at a radial position $r\leq a$ is
\begin{align}
    U =&\; - \frac{G M m}{R} \Big(\frac{3}{2} - \frac{r^2}{2\;\!R^2}\Big)\,, & (r\leq a)
\end{align}
and the (instantaneous) kinetic energy at a radial position $r \leq a$ is
\begin{align}
    K =&\; E - U = \frac{G M m}{2\;\!R^3}\big(a^2-r^2) \,. & (r\leq a)
\end{align}
Averaging over radial positions, the mean kinetic energy is 
\begin{align}\label{eq:mean-K-th2}
    \langle K \rangle = \frac{1}{a}\int_0^a K\;\! dr = \frac{G M m \;\! a^2}{3\;\!R^3}
\end{align}
As long as $v \geq v_*$, DM-nuclear encounters are dominated by DM movements, thus the mean scattering timescale is
\begin{align}
    \langle \Delta t\rangle = (n_*\;\!\sigma_A\;\! v)^{-1}
\end{align}
where $v = \sqrt{2\;\!\langle K\rangle/m}$. From Eq.~\eqref{eq:E-th2} we have
\begin{align}
    \frac{da}{dE}  = \frac{R^3}{G M m \;\!a}
\end{align}
and using treating scattering again as a continuous process
\begin{align}
    \frac{da}{dt} = \frac{da}{dE}\;\!\frac{dE}{dt} = - A_2\;\!\big(B_2\;\!a^2 - 1\big)
\end{align}
with
\begin{align}
    A_2 = &\; \sqrt{\frac{2\;\!R^3}{3\;\!G M}}\frac{n_*\;\!\sigma_A\;\!\beta_+\;\!\langle K_*\rangle}{4m}\,,\\
    B_2 =&\; \frac{G M m}{3 R^3\;\!\langle K_*\rangle}
\end{align}
Integrating, we have
\begin{align}
    t_2 = \int\! dt  = -\frac{1}{A_2}\int_R^{R_2}\!\!\! \frac{da}{B_2\;\!a^2-1} 
\end{align}
where $R_2$ is given by $v=v_*$, or $\sqrt{2\;\! \langle K\rangle/m} = \sqrt{2\;\!\langle K_*\rangle/m_*}$. Using Eq.~\eqref{eq:mean-K-th2} with $a=R_2$ and solving for $R_2$, we have
\begin{align}
    R_2 = \Big(\frac{9\;\!k_{\rm B}\;\!T_{*\infty}\;\!R^3}{2\;\!G M m_*}\Big)^{\!1/2}
\end{align}
If we pose $x=\sqrt{B_2}\;\!a$, then $x \in \sqrt{m/m_*}[1,R/R_2]$ is larger than $1$ and we have
\begin{align}
    \int \frac{dx}{1-x^2} = &\; \frac{1}{2}\ln\Big|\frac{1+x}{1-x}\Big| = {\rm arcoth}\, x \,, & (x>1)
\end{align}
and
\begin{align}
 t_2 = \int\! dt = \frac{1}{A_2\;\!\sqrt{B_2}}\Big[{\rm arcoth}\, x\Big]_{\sqrt{B_2}R}^{\sqrt{B_2}R_2}
\end{align}

We have
\begin{align}
    t_2 = \frac{1}{2\;\!A_2\;\!\sqrt{B_2}}\ln \Big[\frac{(\sqrt{B_2}\;\!R_2+1)(\sqrt{B_2}\;\!R-1)}{(\sqrt{B_2}\;\!R_2-1)(\sqrt{B_2}\;\!R+1)}\Big]
\end{align}
Using $\ln(1+x) \simeq x-x^2$ and $\ln(1-x) \simeq -x-x^2$ where $x\simeq 0$, 

\begin{align}
t_2 = \Big(\frac{18\;\!m}{n_*^2\;\!\sigma_A^2\;\!\beta_+^2\;\!\langle K_*\rangle}\Big)^{\!1/2}\;\! \Bigg[2\sqrt{\frac{m_*}{m}} - 2 \sqrt{\frac{3\;\!R\;\!\langle K_*\rangle}{G M m}} \Bigg]
\end{align}
or, assuming $m_* \ll m$, 
\begin{align}
    t_2 \simeq \frac{3\;\!m}{2\;\!\rho_*\;\!\sigma_A\;\!v_*}\Big(1-\sqrt{3}\frac{v_*}{v_e}\Big)
\end{align}
where $v_e = \sqrt{2\;\!G M/R}$.

\subsection{Third stage}
During this stage, scatterings are dominated by ion movements. Therefore, the mean scattering timescale is
\begin{align}
    \langle \Delta t\rangle = (n_*\;\! \sigma_A\;\! v_*)^{-1}\,,
\end{align}
and we have
\begin{align}
    t_3 = \int\!dt = \frac{1}{A_3}\int_{R_2}^{R_{\rm th}} \frac{a\;\!da}{1-B_3\;\!a^2}\,.
\end{align}
The limits are $\sqrt{B_3}R_{\rm th} = 1$ and $\sqrt{B_3}\;\! R_2 = \sqrt{m/m_*}$ and the time is parametrically infinite
\begin{align}
    t_3 = \frac{1}{A_3\;\! B_3}\int_1^{\sqrt{m/m_*}} \frac{x\;\! dx}{x^2-1} = \frac{1}{A_3\;\! B_3}\Bigg[\ln(x^2-1)\Bigg]^{\sqrt{m/m_*}}_1\,.
\end{align}
However, we can consider that the DM particle is thermalized when it reaches the thermal energy of stellar particles within $\langle\delta\Delta E\rangle$, where $\delta E$ is the root mean square energy transfer,
\begin{align}
    \langle \delta \Delta E \rangle = \sqrt{\langle \Delta E^2\rangle - \langle \Delta E\rangle^2} = \frac{\beta_+}{4\sqrt{2}} \sqrt{K^2+K_*^2}\,.
\end{align}
Then we integrate $t_3$ to the radius $R_3$, defined by $K = K_*+\delta \Delta K$, and, solving for $K$, we find
\begin{align}
    K = K_*\Big[1+\sqrt{1-(1-\beta_+/16\sqrt{2})^2}\Big(1-\beta_+^2/16\sqrt{2}\Big)^{\!-1/2}\Big]\,.
\end{align}
For $\beta_+ \ll 1$, 
\begin{align}
    K \simeq K_*\Big(1+2^{5/4}\frac{m_*}{m}\Big)\,.
\end{align}
Then we have 
\begin{align}
    R_3^2 = \frac{3\;\!R^3\langle K_*\rangle}{G M m} \Big(1+2^{5/4}\frac{m_*}{m}\Big)\,,
\end{align}
and
\begin{align}
    \sqrt{B_3}\;\! R_3 \simeq 1 + 2^{1/4}\frac{m_*}{m}\,,
\end{align}
and the integral is finite
\begin{align}
    t_3 =&\; \frac{1}{A_3\;\!B_3}\Bigg[ \ln\Big(\frac{m}{m_*}-1\Big) - \ln\Big(1 + 2^{1/4}\frac{m_*}{m} -1\Big)\Bigg] \nonumber \\
    &\; \simeq \frac{2}{A_3\;\!B_3}\;\!\ln \Big(\frac{m}{m_*}\Big)\,.
\end{align}
Introducing the values of $A_3$ and $B_3$,
\begin{align}
    t_3 = \frac{3\;\!m}{n_*\;\!\sigma_A\;\!v_*\;\!m_*}\;\!\ln\frac{m}{m_*}\,.
\end{align}
Finally, summing up, the total thermalization timescale is
\begin{align}
    t_{\rm th} = t_1+t_2+t_3 = \frac{3\;\!m}{\rho_*\;\!\sigma_A\;\!v_*}\Big[\frac{3\sqrt{2}\pi}{16}\frac{v_*}{v_{\rm gal}} + \frac{1}{2} + \ln\Big(\frac{m}{m_*}\Big)\Big]\,.
\end{align}

\section{SCATTERING ENERGY TRANSFER}

\subsection{Classical regime}\label{sec:Delta-E}
Consider scattering of two rigid body spheres with mass $m$ and $m_*$ and initial scalar velocities $v$ and $v_*$ respectively. A general scattering event occurs in a plane. In full generality, we can choose a coordinate system such that the line of centers coincides with the $x$ axis (in other words, the contact angle is zero). Momentum conservation along the line of centers and kinetic energy conservation imply
\begin{align}
    & m\;\! v \cos\theta + m_*\;\! v_* \cos\theta_* = m\;\! v_x' + m_*\;\! v_{*x}'\,, \\
    & m\;\! v^2 + m_*\;\! v_*^2 = m\;\! v'^2+m_*\;\! v_*'^2\,,
\end{align}
where $\theta$ and $\theta_*$ are the precollision movement angles (with respect to the $x$ axis) of $m$ and $m_*$, respectively, and primes indicate post-collision quantities.
Momentum perpendicular to the line of centers is conserved for each mass, so we obtain immediately,
\begin{align}
    & v_y' = v\sin\theta \,,\\
    & v_{*y}' = v_* \sin \theta_*\,,
\end{align}
With four unknowns ($v_x',\,v_y',\,v_{*x}',\,v_{*y}'$) and four equations, the system has a unique nontrivial solution,
\begin{align}
    v_x' =&\; \frac{m-m_*}{m+m_*}\;\! v \cos\theta + \frac{2\;\! m_*}{m+m_*}\;\! v_* \cos\theta_* \,,
\end{align}
Using $(m-m_*)^2/(m+m_*)^2=1-4\;\!m\;\!m_*/(m+m_*)^2$ and defining $\beta_+ \equiv 4\;\!m\;\!m_*/(m+m_*)^2$, we can express the total post-shock velocity squared as
\begin{align}
    v'^2 =&\; v^2  + \beta_+\Big[\frac{m_*}{m}\;\! {v_*}^{\!2} \cos^2\!\theta_* - \;\! v^2 \cos^2\!\theta \nonumber \\
    &\; + \frac{m\!-\!m_*}{m}\;\!v\;\!v_* \cos\theta\;\!\cos\theta_*\Big] \,.
\end{align}
In terms of energy (for $E=K$)
\begin{align}\label{eq:Delta-E-non-rel}
    \Delta E =&\; E'-E =   \beta_+\;\! \Big[E_* \cos^2\!\theta_*  - E\cos^2\!\theta \nonumber \\ 
    &\; + (m\!-\!m_*)\sqrt{E\;\!E_*/m\;\!m_*}\;\! \cos\theta\;\!\cos\theta_*\Big]\,.
\end{align}
Since all precollision movement angles are equally likely, the average energy transfer is
\begin{align}\label{eq:DeltaE-WD}
    \langle \Delta K\rangle = \frac{1}{(2\;\!\pi)^2}\!\int_0^{2\;\!\pi}\!\!\!\int_0^{2\;\!\pi}\!\!\!\Delta E \;\!d\theta\;\!d\theta_* =  \frac{\beta_+}{2}\big(K_*-K\big)\,.
\end{align}
In the main text, we omit the brackets and average additionally over velocity distribution.
Note that, whenever both species are at thermal equilibrium, \eqref{eq:DeltaE-WD} implies that the energy transfer is zero.

\subsection{Relativistic regime}\label{sec:Delta-E-relativistic}
In the relativistic case, we have momentum conservation along the line of centers 
\begin{align}\label{eq:rel-1}
    p \cos\theta + p_* \cos\theta_* = p_x' + p_{*x}'\,,
\end{align}
and energy conservation
\begin{align}\label{eq:rel-2}
    E+E_* = E'+E_*'\,.
\end{align}
Momentum conservation perpendicular to the line of centers for each mass 
\begin{align}
    &\; p_y' = p \sin\theta\,, \label{eq:rel-3}\\
    &\; p_{*y}' = p_* \sin\theta_* \,.\label{eq:rel-4}
\end{align}
First, we eliminate $p_*'$. On the one hand, we isolate the $x$ and $y$ components using Eqs.~\eqref{eq:rel-1} and \eqref{eq:rel-4} and writing the sum of squares
\begin{align}
    p_*'^2 = p_{*x}'^2 + p_{*y}'^2 = &\; p^2 \cos^2\theta + 2\;\!p\;\!p_* \cos\theta\;\! \cos\theta_* + p_*^2 + p_x'^2 \nonumber \\
    &\; - 2\;\!(p\cos\theta + p_* \cos\theta_*)\;\!p_x' \label{eq:rel-5}
\end{align}
On the other hand, from Eq.~\eqref{eq:rel-2} we have
\begin{align}
    p_*'^2 = &\; E_*'^2 - m_*^2 = (E+E_*-E')^2 - m_*^2 \nonumber \\
     &\; E^2 + p_*^2 + E'^2 + 2\;\!E\;\!E_* - 2\;\!(E+E_*)\;\!E'\,. \label{eq:rel-6}
\end{align}
Equating \eqref{eq:rel-5} and \eqref{eq:rel-6} and using Eq.~ \eqref{eq:rel-3} in the form $p_x'^2 = p'^2 - p_y'^2 = p'^2 - p^2 \sin^2\theta$, we obtain after some algebra
\begin{align}
    &\;\big[(E+E_*)^2 - C^2\big]\;\! E'^2 - 2\;\!B\;\!(E+E_*)\;\!E' \nonumber \\
    &\;+ \big[B^2 + C^2\;\!(m^2+p^2 \sin^2\theta)\big] = 0\,,
\end{align}
where we have defined 
\begin{align}
    &\; B \equiv m^2 + E\;\!E_* + p^2 \sin^2\theta - p\;\!p_* \cos\theta\;\! \cos\theta_* \,,\\
    &\; C \equiv p \cos\theta + p_* \cos\theta_* \,.
\end{align}
Solving for $E'$, we have
\begin{align}
    \big(E'\big)_{\pm} = \frac{B\;\!(E+E_*) \pm C\;\!\sqrt{|\Delta|}}{(E+E_*)^2-C^2}\,,
\end{align}
where the discriminant is
\begin{align}
    \Delta = B^2 - (m^2+p^2\sin^2\theta)[(E+E_*)^2-C^2]\,.
\end{align}
The difference is
\begin{align}
    \big(\Delta E\big)_{\pm} = E'-E = \frac{B\;\!(E+E_*)- E\;\![(E+E_*)^2-C^2] \pm C\;\!\sqrt{|\Delta|}}{(E+E_*)^2-C^2}\,.
\end{align}
It can be shown, after long algebra, that 
\begin{align}
    C\;\!\sqrt{|\Delta|} = B\;\!(E+E_*)- E\;\![(E+E_*)^2-C^2]\,.
\end{align}
Therefore, the nontrivial solution is the plus solution. After some simplification, we obtain
\begin{align}\label{eq:Delta-E-rel-app}
    \Delta E = \frac{2\;\![E p_*^2 \cos^2\!\theta_* + (E\!-\!E_*)\;\!p\;\!p_* \cos\theta\;\!\cos\theta_* - E_*\;\!p^2 \cos^2\!\theta]}{(E+E_*)^2-(p\cos\theta+p_*\cos\theta_*)^2}\,.
\end{align}
It is easy to verify that for $m\gg p$ and $m_*\gg p_*$, Eq.~\eqref{eq:Delta-E-rel-app} reduces to the non-relativistic Eq.~\eqref{eq:Delta-E-non-rel}.

\section{COLLISION RATE}\label{sec:collision-rate}
Here we assume that the cross section does not vary with the relative velocity. The characteristic collision time is
\begin{align}\label{eq:Delta-t-mean}
    \Delta t = \frac{1}{n_*\;\!\sigma\;\! \langle v_{\rm rel}\rangle}\,,
\end{align}
where $v_{\rm rel}$ is the relative velocity between colliding particles \cite{1975ctf..book.....L}
\begin{align}
    v_{\rm rel} = &\;  \frac{\sqrt{(\bs{v}-\bs{v}_*)^2-(\bs{v}\times \bs{v}_*)^2}}{1-\bs{v} \cdot \bs{v}_*} 
     =  \frac{\sqrt{(p\cdot p_*)^2-m^2\;\!{m_*}^{\!2}}}{p\cdot p_*}\,,
\end{align}
and the mean $\langle\ldots\rangle$ is taken over the J\"{u}ttlich distribution (relativistic generalization of Maxwell distribution)
\begin{align}
    f_{\rm J}(\bs{p}) = (4\;\!\pi\;\!m^2\;\!T\;\!K_2(x))^{-1}\;\!\exp[-\sqrt{\bs{p}^2+m^2}/T]\,.
\end{align}
The mean can be shown to be (see, for example, \cite{2017IJMPA..3230002C})
\begin{align}\label{eq:vrel-mean-app}
    \langle v_{\rm rel}\rangle = \frac{2\;\![(1+\zeta)^2\;\!K_3(\xi)-(\zeta^2-1)\;\!K_1(\xi)]}{\xi\;\! K_2(x)\;\!K_2(x_*)}\,,
\end{align}
where $\xi = x+x_*$, $\zeta = (x^2+x_*^2)/2\;\!x\;\!x_*$ are auxiliary variables and $x=m\;\!c^2/k_{\rm B}\;\!T$ and $x_*=m_*\;\!c^2/k_{\rm B}\;\!T_*$ are standard thermal variables, $K_i(x)$ is the modified (or hyperbolic) Bessel function of the second kind (not to confuse with the kinetic energy that we denote $K$ as well). 

For $n$ an integer, the modified Bessel functions of the first and second kind are
\begin{align}
    I_n(x) = 
    K_n(x) = \lim_{\alpha\to n}\frac{\pi}{2}\frac{I_{-\alpha}(x)-I_{\alpha}(x)}{\sin\alpha\;\!\pi} \,,
\end{align}
where $\alpha$ is a noninteger. 

The following asymptotic formula for large arguments is useful
\begin{align}
    K_{\nu}(x) =&\; \Big(\frac{\pi}{2\;\!x}\Big)^{1/2}\;\!e^{-x}\;\!\Big[1 + \frac{1\!-\!4\;\!\nu^2}{8\;\!x}+\frac{9\!-\!40\;\!\nu^2\!+\!16\;\!\nu^4}{128\;\!x^2} \nonumber \\
    &\; + \mathcal{O}(x^{-3})\Big]\,.
\end{align}
For numerical purposes, it is useful to rewrite the following form
\begin{align}\label{eq:vrel-mean-tilde}
    \langle v_{\rm rel}\rangle = \Big(\frac{8\;\!x\;\!x_*}{\pi\;\!\xi}\Big)^{\!1/2}\/\! \frac{(1\!+\!\zeta)^2\;\!\tilde{K}_3(\xi)-(\zeta^2\!-\!1)\;\!\tilde{K}_1(\xi)}{\xi\;\! \tilde{K}_2(x)\;\!\tilde{K}_2(x_*)}\,,
\end{align}
where we have defined
\begin{align}
    \tilde{K}_i(y) = &\; K_i(y)\;\!\sqrt{\frac{2\;\!y}{\pi}}\;\!e^y \,.
\end{align}
The advantage is that this has a simple expansion in the nonrelativistic limit, $y\to \infty$ (which we use as soon as $y>100$)  
\begin{align}\label{eq:K-tilde-i-y}
    \tilde{K}_i(y) = 1 + \frac{4\;\!i^2\!-\!1}{8\;\!y}+\frac{16\;\!i^4\!-\!40\;\!i^2\!+\!9}{128\;\!y^2} + \mathcal{O}(y^{-3})\,.
\end{align}
Another useful limit is when $\zeta \gg 1$ (typically $\zeta > 10^{12}$, or machine precision), corresponding to either $x\gg x_*$ or $x \ll x_*$. Then we have
\begin{align}\label{eq:vrel-mean-tilde-zeta}
    \lim_{\zeta\to\infty}\langle v_{\rm rel}\rangle \simeq \Big(\frac{8\;\!x\;\!x_*}{\pi\;\!\xi}\Big)^{\!1/2}\;\! \frac{2\;\!\zeta \tilde{K}_3(\xi)+\zeta^2\;\!\Delta\tilde{K}_{31}(\xi)}{\xi\;\! \tilde{K}_2(x)\;\!\tilde{K}_2(x_*)}\,,
\end{align}
where 
\begin{align}
    \Delta\tilde{K}_{ij}(y) = \frac{4\;\!(i^2\!-\!j^2)}{8\;\!y} + \frac{16\;\!(i^4\!-\!j^4)-40\;\!(i^2\!-\!j^2)}{128\;\!y^2} + \mathcal{O}(y^{-3}) \,.
\end{align}
Note that in Eq.~\eqref{eq:Delta-t-mean} assumes DM particles remain nonrelativistic. 
Comparing Eqs.~\eqref{eq:vrel-mean-tilde} and \eqref{eq:K-tilde-i-y}, it is easy to verify that for both $x\to \infty$ and $x_*\to \infty$, the well-known nonrelativistic expression is recovered
\begin{align}\label{eq:vrel-mean-nonrel}
    \langle v_{\rm rel}\rangle = \Big[\frac{8\;\!(x+x_*)}{\pi\;\! x\;\!x_*}\Big]^{1/2}\,.
\end{align}

\section{DETAILS ON THE MERGER DELAY}\label{sec:details-merger-rate}
Using the binary population synthesis code \textsc{StarTrack}, Ref.~\cite{2010Natur.463...61P} computes the rate of double degenerate mergers with primary mass between 0.85 and 1.05~$M_{\odot}$ and secondary with mass ratio $0.9<M_2/M_1<1.0$. These systems count on contributions from three distinct evolutionary channels:

(a) Prompt ($<0.1$~Gyr) delay times originate from 6.0--7.5~$M_{\odot}$ zero-age main sequence masses (rare) and close initial orbits ($a_0 < 200~R_{\odot}$)\footnote{This must be an error. In order to obtain (b)/(a)~$=3$ as stated by the author, we must have $a_0 \sim 60$--200$R_{\odot}$ instead of {$a_0<200R_{\odot}$}.}, undergo two common envelopes, and comprise 25\% of the channels (a) and (b) together.

(b) Intermediate (1--3~Gyr) delay times originate from 4.8--5.8~$M_{\odot}$ zero-age main sequence masses and wider initial separations ($a_0 \sim 80$--$1000~R_{\odot}$), undergo only one common envelope, and comprise 75\% of channels (a) and (b) together.

(c) Very long ($\sim 10$~Gyr) delay times consist of binaries with zero-age main sequence component masses in the range 3.8--4.5~$M_{\odot}$ and large spread in initial separations $a_0 \sim 100$--$2000~R_{\odot}$. These experience only one common envelope (when the primary has already evolved into a WD), and the mass-losing star is a bloated late-AGB star. The orbital separation upon ejection of the common envelope is $a\sim 3~R_{\odot}$ implying a multi-Gyr delay time.

For our purpose only in channel (c) is of interest but its incidence has not been specified by Ref.~\cite{2010Natur.463...61P}. Assuming a \citet{1955ApJ...121..161S} initial mass function, $N(M)\;\!dM \propto M^{-2.35}dM$, 
and logarithmically uniform initial period distribution  \citep[e.g.][]{2015MNRAS.447.1713B},
We find that the respective fractions of the channels (a), (b), and (c) are 10\%, 30\%, and 60\%. We assume that these respective fractions persist the same when considering slightly higher component masses.

\end{document}